\newif\ifconfver
\confverfalse      %declaring conference version false
\confvertrue        %declaring conference version true

\newif\ifplainver  %declare a plain version
\plainvertrue
\ifplainver
    \confverfalse   %automatically disable conf. version argument if it's plain
\fi

\ifconfver
     \documentclass[10pt,twocolumn,twoside]{IEEEtran}
\else
    \ifplainver
        \documentclass[11pt]{article}
        \usepackage{fullpage}
    \else
        \documentclass[12pt,draftcls,onecolumn]{IEEEtran}
    \fi
\fi

\usepackage{calc,amsfonts,amssymb,amsmath,bm,url,color,theorem,graphicx,cite}
\usepackage{enumerate}
\usepackage{psfrag,float}
\usepackage{algorithm}
\usepackage{array}
\usepackage{algorithmic}
\usepackage{subcaption,epstopdf}
\usepackage{ragged2e}
\definecolor{orange}{RGB}{255,107,0}

\newtheorem{Fact}{Fact}
\newtheorem{Lemma}{Lemma}
\newtheorem{Prop}{Proposition}
\newtheorem{Theorem}{Theorem}

\newtheorem{Corollary}{Corollary}

\theorembodyfont{\rmfamily}

\newtheorem{Remark}{Remark}

\newcommand\bx{\ensuremath{{\bm x}}}
\newcommand\by{\ensuremath{{\bm y}}}

\newcommand\bh{\ensuremath{{\bm h}}}
\newcommand\bH{\ensuremath{{\bm H}}}

\newcommand\bz{\ensuremath{{\bm z}}}

\newcommand\bX{\ensuremath{{\bm X}}}
\newcommand\bZ{\ensuremath{{\bm Z}}}

\newcommand\bc{\ensuremath{{\bm c}}}
\newcommand\ba{\ensuremath{{\bm a}}}

\newcommand\bb{\ensuremath{{\bm b}}}

\newcommand\bd{\ensuremath{{\bm d}}}

\newcommand\bu{\ensuremath{{\bm u}}}
\newcommand\bv{\ensuremath{{\bm v}}}

\newcommand\bU{\ensuremath{{\bm U}}}

\newcommand\bs{\ensuremath{{\bm s}}}

\newcommand{\Rbb}{\mathbb{R}}
\newcommand{\Cbb}{\mathbb{C}}

\newcommand{\setX}{\mathcal{X}}

\newcommand{\setU}{\mathcal{U}}

\newcommand{\setS}{\mathcal{S}}

\newcommand{\setD}{\mathcal{D}}

\newcommand{\setP}{\mathcal{P}}

\newcommand{\ind}{I}

\newcommand{\Exp}{\mathbb{E}}
\newcommand{\jj}{\mathfrak{j}}
\newcommand{\dec}{\mathrm{dec}}

\newcommand{\bzero}{{\bm 0}}

\newcommand\conv{\ensuremath{{\rm conv  \,}}}

\newcommand\dRi{\ensuremath{{d_i^R}}}
\newcommand\dIi{\ensuremath{{d_i^I}}}
\newcommand\SEPR{\ensuremath{{{\sf SEP}^R_{i,t}}}}
\newcommand\SEPI{\ensuremath{{{\sf SEP}^I_{i,t}}}}
\newcommand\aRi{\ensuremath{{a_{i,t}^R}}}
\newcommand\aIi{\ensuremath{{a_{i,t}^I}}}
\newcommand\bRi{\ensuremath{{b_{i,t}^R}}}
\newcommand\bIi{\ensuremath{{b_{i,t}^I}}}
\newcommand\cRi{\ensuremath{{c_{i,t}^R}}}
\newcommand\cIi{\ensuremath{{c_{i,t}^I}}}
\newcommand{\setbarU}{\bar{\mathcal{U}}}

\newcolumntype{M}[1]{>{\centering\arraybackslash}m{#1}}

\makeatletter
\def\bstctlcite{\@ifnextchar[{\@bstctlcite}{\@bstctlcite[@auxout]}}
\def\@bstctlcite[#1]#2{\@bsphack
  \@for\@citeb:=#2\do{%
    \edef\@citeb{\expandafter\@firstofone\@citeb}%
    \if@filesw\immediate\write\csname #1\endcsname{\string\citation{\@citeb}}\fi}%
  \@esphack}
\makeatother
\begin{document}

\bibliographystyle{IEEEtran}

%--- I do things quite strangely here to accommodate three style modes.
%--- input title and abstract here; it applies to all modes
%--- it's too complex to do authors or they are input for each mode
\newcommand{\papertitle}{
A Framework for One-Bit and Constant-Envelope Precoding over Multiuser Massive MISO Channels}

\newcommand{\paperabstract}{
Consider the following problem:
A multi-antenna base station (BS) sends multiple symbol streams to multiple single-antenna users via precoding.
%, assuming perfect channel information at the BS.
However, unlike conventional multiuser %beamforming,
precoding,
the transmitted signals are subjected to binary, unit-modulus, or even discrete unit-modulus constraints.
Such constraints arise in the one-bit and constant-envelope (CE) massive MIMO scenarios,
wherein high-resolution digital-to-analog converters (DACs) are replaced by one-bit DACs and phase shifters, respectively, for cutting down hardware cost and energy consumption.
%One-bit and CE multiuser precoding poses difficult design problems owing to the non-convexity of their constraints.
Multiuser precoding under one-bit and CE restrictions poses significant design difficulty.
In this paper we establish a framework for designing multiuser precoding under the one-bit, continuous CE and discrete CE scenarios---all within one theme.
We first formulate a precoding design that focuses on minimization of the symbol-error probabilities (SEPs), assuming quadrature amplitude modulation (QAM) symbol constellations.
%Currently available designs either consider other formulations, such as minimum mean-square error, or they are related to SEP in an indirect or case-specific manner.
We then devise an algorithm for our SEP-based design.
The algorithm combines i) a novel penalty method for handling binary, unit-modulus and discrete unit-modulus constraints; and
ii) a first-order non-convex optimization recipe custom-built for the design.
Specifically, the latter is an inexact majorization-minimization method via accelerated projected gradient,
which, as shown by simulations, runs very fast and can handle a large number of decision variables.
Simulation results indicate that the proposed design offers significantly better bit-error rate performance than the existing designs.
}

%--------

\ifplainver

    %\date{May 30, 2014}

    \title{\papertitle}

    \author{
    Mingjie Shao$^\dag$, Qiang Li$^{\ddag}$, Wing-Kin Ma$^\dag$, and Anthony Man-Cho So$^\star$ \\ ~ \\
    $^\dag$Department of Electronic Engineering, The Chinese University of Hong Kong, \\
    Hong Kong SAR of China \\ ~ \\
    $^\ddag$School of Information and Communications Engineering, \\
    University of Electronic Science and Technology of China, China \\ ~ \\
    $^\star$Department of Systems Engineering and Engineering Management, \\ The Chinese University of Hong Kong,
    Hong Kong SAR of China \\ ~ \\
    E-mails: mjshao@ee.cuhk.edu.hk, lq@uestc.edu.cn, wkma@ee.cuhk.edu.hk, \\
    manchoso@se.cuhk.edu.hk
    }

    \maketitle

    \begin{abstract}
    \paperabstract
    \end{abstract}
%    \begin{keywords}\vspace{-0.0cm}
%        massive MIMO, multiuser precoding, one-bit, constant envelope,  penalty method
%    \end{keywords}

\else
    \title{\papertitle}

    \ifconfver \else {\linespread{1.1} \rm \fi

    \author{Mingjie Shao, Qiang Li, Wing-Kin Ma, and Anthony Man-Cho So}

    \maketitle

    \ifconfver \else
        \begin{center} \vspace*{-2\baselineskip}
        %11th Revision, \today \\[2\baselineskip]
        \end{center}
    \fi

    \begin{abstract}
    \paperabstract
    \\\\
    \end{abstract}

    \begin{IEEEkeywords}\vspace{-0.0cm}
        massive MIMO, multiuser precoding, one-bit, constant envelope,  penalty method
    \end{IEEEkeywords}

%    \begin{keywords}\vspace{-0.0cm}
%        ...
%    \end{keywords}

    \ifconfver \else \IEEEpeerreviewmaketitle} \fi

 \fi

\ifconfver \else
    \ifplainver
    		\newpage
    	\else
        	\newpage
\fi \fi
%---------------------------------------------------------------------------
\section{Introduction}

\justify
Lately, there has been great enthusiasm for researching
%one-bit
coarsely quantized
and constant-envelope (CE) techniques for massive multiple-input multiple-output (MIMO) systems.
It has been widely recognized that massive MIMO provides many benefits such as enhanced spectral efficiency and massive connectivity, $\!\!$
but $\!\!$ it $\!\!$ is $\!\!$ also $\!\!$ known $\!\!$ that $\!\!$ the $\!\!$ number $\!\!$ of $\!\!$ analog-to-digital converters (ADCs)/digital-to-analog converters (DACs) and radio-frequency (RF) front ends needs to scale by the same number as the very large number of antennas in massive MIMO---which introduces significant issues with hardware cost and energy consumption.
The study of
%one-bit
coarsely quantized
and CE techniques is motivated by the need to overcome such issues.

One direction to deal with the ADC/DAC-cost issues is to simply replace the currently-used high-resolution ADCs/DACs by $\!\!$ low-resolution $\!\!$ ones, $\!\!$ particularly, $\!\!$ the $\!\!$ very $\!\!$ cheap $\!\!$ one-bit $\!\!$ ADCs/DACs.
For massive MIMO uplink, it has been demonstrated that MIMO detection with one-bit ADCs can actually achieve promising performance \cite{Choi2015,Choi2016,Studer2016}.
For massive MIMO downlink, MIMO precoding with one-bit DACs, or simply one-bit precoding, is a relatively new problem.
There is an additional reason for considering one-bit precoding.
RF power amplifiers (PAs) are known to waste a significant portion of energy when they are operated under high power back-off mode for providing linear amplification of high peak-to-average power ratio (PAPR) signals.
A popular way to mitigate this issue is to employ pre-distortion \cite{ghannouchi2009behavioral},
but pre-distortion also raises hardware requirement on a per-antenna scale.
On the other hand, if we transmit CE signals, then PAs can be operated under low back-off and can have high power efficiency.
It happens that one-bit precoding restricts the transmitted signal of each antenna to be of CE, specifically, in a $4$-ary phase shift keying (PSK) form.
Thus, one-bit precoding provides an opportunity to substantially cut down energy consumption and also hardware complexity associated with PAs.

One-bit precoding is not the only CE signaling strategy for encouraging use of inexpensive and energy-efficient PAs.
Another strategy is to replace the high-resolution DACs with constant-amplitude analog phase shifters. This is known as {\em CE precoding} in the literature.\footnote{ One-bit precoding is also a constant envelope scheme {\em per se}, but following the convention in the literature we will use ``CE precoding'' to refer to the phase shifter-based CE approach only.}
In CE precoding, the transmitted signal of each antenna is restricted to take an $M$-ary PSK form.
Or, if the phase resolution is high enough, we may assume the signal to take a continuous constant modulus form.
From the precoding design viewpoint, one-bit precoding can be regarded as a special case of CE precoding where $M=4$.

%CE precoding, which works by replacing high-resolution DACs by analog phase shifters, was motivated by exactly the same reason.
%It should be mentioned that CE precoding can be regarded as a more general, but also more challenging, version of one-bit precoding, where CE precoding restricts each transmitted signal to take an $M$-ary PSK form or a continuous unit-modulus form.

A very difficult problem that arises, at least at first sight, is how we should design one-bit and CE precoding.
The problem amounts to finding an $M$-ary PSK or constant-modulus transmit signal vector---which is generally hard to manipulate algebraically---such that receivers will receive their symbol streams with minimal distortions.
Many concepts we know in conventional precoding, which, loosely speaking, consider the transmit signal vector lying in the free space, do not apply when the binary, $M$-ary PSK or unit-modulus restrictions set in.
Despite such difficulty, one-bit and CE precoding designs have triggered much interest most recently.
The results in the current literature, due to the emerging nature of the problem, are somewhat scattered and not well unified;
e.g., they may specialize in a particular scenario (e.g., only one-bit or CE)
%a specific design criterion,
and/or a specific symbol constellation (e.g., only PSK).
Here we attempt to taxonomize the various design methods.
But before we proceed, we should mention that one-bit and CE precoding for the single-user multiple-input single-output (MISO) scenario has been well-studied \cite{Mohammed2012,Pan2014,Sohrabi2018},
and the multiuser MISO scenario will be our focus.
\begin{enumerate}[1.]
	\item {\em Quantized Linear Precoding:} \
	The idea is to apply quantization, such as one-bit quantization in one-bit precoding, to a ``free-space'' linear precoder output, such as zero-forcing.
	Such quantized linear precoding is natural and simple to implement,
	but its performance is not as competitive as that of the approaches to be described next.
	Some studies analyze the performance of quantized linear precoding~\cite{Mezghani2009,Saxena2017}, which is useful in understanding the performance gap
	%between one-bit and free-space precoding;
	before and after quantization;
	some considers improved designs via symbol perturbations~\cite{Swindlehurst2017}.
	
	\item {\em Distortion Minimization:} \
	This approach designs the transmit signal vector directly, rather than quantizing a free-space precoder output.
	This requires us to solve an optimization problem, but better symbol-error probability performance has been observed compared to quantized linear precoding.
	The rationale is to minimize distortions that appear in the received signals, relative to the ground-truth symbols.
	Criteria used in the literature include minimum mean square error (MMSE)~\cite{Jacobsson2017} and multiuser interference minimization in the least squares sense~\cite{Mohammed2013}; see \cite{Jacobsson2016,Jacobsson2017a,Jacobsson2018,wang2018finite} and \cite{Mohammed2013a, Chen2014,Liu2017,Kazemi2017} for more.
	In one-bit precoding,
	the MMSE-based algorithm in \cite{Jacobsson2017}, called SQUID, is particularly popular.
	
	\item {\em Constellation-Dependent Designs:} \
	The previously described distortion-minimization methods use second-order metrics to measure distortion, which do not consider symbol constellations.
	It has been recently known that, even for free-space precoding, distortion or interference can be beneficially aligned to improve symbol-error probability (SEP) performance when symbol constellation structures are taken into design consideration \cite{masouros2009dynamic,alodeh2018symbol}.
	Such notion is known as constructive interference or symbol-level precoding.
	Some recent works begin to exploit specific symbol constellations in one-bit and CE precoding \cite{Amadori2017,Jedda2017,Jedda2018}.
	% by trying to push the received signal points away from the symbol decision boundary.
	A few most recent works take a more systematic approach by working on SEP directly \cite{shao2018,Sohrabi2018,shao2018CE,shao2018PSK}.
	Like the previous distortion-minimization approach, constellation-dependent designs require optimization.
	It has been illustrated that constellation-dependent designs can provide further improved SEP performance compared to the previous two approaches.
\end{enumerate}

The above taxonomy is based on design formulations.
The next challenge is with the optimization of the subsequent design problem, which, as mentioned, is hard owing to the discrete and/or non-convex equality constraints with one-bit and CE precoding.
To make the matter even more complex, the current algorithmic developments are intimately linked with factors such as the design formulation chosen, the scenario (e.g., one-bit, or CE?), and the symbol constellation used.
Simply speaking, some works use convex relaxation, and some combinatorial optimization.

\subsection{This Work and Contributions}

In this paper, we propose a framework for one-bit and CE precoding under the multiuser MISO downlink scenario.
We consider a minimax SEP design formulation, with an emphasis on developing efficient optimization methods to tackle the formulation.
Our framework is constellation-dependent and is built for the QAM  constellation.
%It should be noted that
Our framework can be directly applied to the $M$-ary PSK  constellation by applying the optimization methods in this paper to the $M$-ary PSK formulation we studied in \cite{shao2018PSK},
although this direction will not be described owing to page limitation.
As will be shown by simulation results, the proposed framework outperforms the existing designs in terms of SEP performance.
The key contributions of this paper are summarized as follows.
\begin{enumerate}[1.]
	\item Few works deal with one-bit and CE precoding in one theme, and this work makes one such endeavor. In particular, our framework can handle discrete CE restrictions, which are difficult and we currently see only a few works that challenge this setting \cite{Kazemi2017,Jedda2018}.
	
	\item Few works tackle SEP directly in their designs.
	A notable work on this direction is the work in~\cite{Sohrabi2018}, which appears concurrently with the conference version of this paper~\cite{shao2018}.
	The work~\cite{Sohrabi2018} focuses more on analyses of one-bit precoding; it also proposed one-bit algorithms based on search heuristics.
	Our work, in comparison, is more toward building an optimization framework for the problem.
	
	\item As a core technical contribution, we establish an optimization method that allows us to transform the design problem, which has discrete and/or non-convex equality constraints, into an optimization problem with convex constraints.
	This method, called the negative square penalty (NSP) method, plays a key role in enabling us to put one-bit, discrete CE, and continuous CE precoding designs in one theme.
	We also custom-build a first-order non-convex optimization algorithm for the transformed problem; it runs very fast and can handle a large number of decision variables, as our simulation results suggest.
	The proposed algorithm is a non-conventional combination of majorization-minimization and accelerated projected gradient, as we will explain.
	
	\item As a more in-depth technical aspect, our framework also designs the QAM inter-point spacings of the users' symbol streams.
	The inter-point spacings are a key factor in enhancing SEP performance.
    Some existing works pre-fix the inter-point spacings \cite{Mohammed2013,Mohammed2013a, Chen2014,Kazemi2017}, some uses analyses to predict \cite{Sohrabi2018}, and some assume identical inter-point spacing for all users \cite{Jacobsson2016,Jacobsson2017a,Jacobsson2018}.
    Our framework jointly optimizes the precoder and the inter-point spacings, and the treatment is more general than the previous.
	
\end{enumerate}

The organization of this paper is as follows.
Section~\ref{sec:background} describes the signal model of one-bit and CE precoding.
Section~\ref{sec:sep} formulates the minimax SEP design problem.
Section~\ref{sec:NSP} develops the NSP method for transforming the design problem, and Section~\ref{sec:gemm} completes the picture by custom-deriving an algorithm for the NSP-transformed design problem.
Simulation results are shown in Section~\ref{sec:simulation}, and we draw the conclusion in Section~\ref{sec:cl}.

\subsection{Notations and Some Basic Notions}
\label{sec:notations}

Our notations are standard; e.g., $\bx$ as a vector, $\bX$ as a matrix, $\setX$ as a set,
the superscripts ``$T$'' and ``$H$'' as the transpose and Hermitian transpose, respectively (resp.), $\Rbb$ and $\Cbb$ as the set of all real and complex numbers, resp., etc.
In addition, $\| \cdot \|_p$ denotes the $\ell_p$ norm, $\| \cdot \|$ simply denotes the Euclidean norm, $\langle \bx, \by \rangle = \Re( \bx^H \by )$ denotes the inner product,
$\conv \setX$ denotes the convex hull of $\setX$,
and
\[
\Pi_\setX (\bx) = \arg\min_{\by \in \setX} \| \bx - \by \|^2
\]
denotes the projection of $\bx$ onto $\setX$.

Some notations and notions concerning optimization are as follows.
Let $\setX \subseteq \Rbb^n$ or $\setX \subseteq \Cbb^n$,
and let $f: \setX \rightarrow \Rbb$.
The gradient of $f$ at $\bx$ is denoted by $\nabla_\bx f(\bx)$ or simply by $\nabla f(\bx)$.
If $\setX \subseteq \Rbb^n$, the definition of the gradient follows the standard definition.
If $\setX \subseteq \Cbb^n$, we define the gradient as
\begin{equation} \label{eq:grad_c}
\nabla_\bx f(\bx) = \nabla_{\Re(\bx)} f(\bx) + \jj \nabla_{\Im(\bx)} f(\bx).
\end{equation}
Consider an optimization problem
\begin{equation} \label{eq:opt_gen}
\min_{\bx \in \setX } f(\bx),
\end{equation}
where $f$ is differentiable, and $\setX$ is non-empty and closed.
Problem~\eqref{eq:opt_gen} can be rewritten as
\begin{equation*}
\min_{\bx } f(\bx) + \ind_\setX(\bx),
\end{equation*}
where $\ind_\setX$ is the indicator function of $\setX$, i.e., $\ind_\setX(\bx) = 0$ for $\bx \in \setX$ and $\ind_\setX(\bx) = \infty$ for $\bx \notin \setX$.
A first-order necessary condition for $\bx$ to be an optimal solution to the above problem, and also Problem~\eqref{eq:opt_gen}, is
\begin{equation} \label{eq:stat_def}
%\langle \nabla f(\bx), \by - \bx   \rangle \geq 0, \quad \forall \by \in \setX.
\bzero \in \nabla f(\bx) + \partial \ind_\setX(\bx),
\end{equation}
where $\partial \ind_\setX(\bx)$ is the limiting subdifferential of $I_\setX$ at $\bx$; see \cite{mordukhovich2006variational} for the definition.
Note that \eqref{eq:stat_def} was established for $\setX \subseteq \Rbb^n$, but one can easily show that the same notion applies to $\setX \subseteq \Cbb^n$ if we adopt the gradient definition in \eqref{eq:grad_c}.
A point $\bx \in \setX$ is said to be stationary if it satisfies \eqref{eq:stat_def}.
By the same vein, a point $\bx \in \setX$ is said to be $\epsilon$-stationary if
\[
{\rm dist}( \bzero, \nabla f(\bx) + \partial \ind_\setX(\bx) ) \leq \epsilon,
\]
where ${\rm dist}(\bx,\setX) = \inf_{\by \in \setX} \| \bx - \by \|$.

We will also need the notion of Lipschitz continuity.
We say $f$ to be Lipschitz continuous on $\setX$ if there exists a constant $L_1 \geq 0$ such that
\[
| f(\bx) - f(\by) | \leq L_1 \| \bx - \by \|, \quad \forall ~ \bx, \by \in \setX,
\]
and the corresponding constant $L_1$ is called a Lipschitz constant of $f$ on $\setX$.
A differentiable $f$ is said to have Lipschitz continuous gradient on $\setX$ if
there exists a constant $L_2 \geq 0$ such that
\[
\| \nabla f(\bx) - \nabla f(\by) | \leq L_2 \| \bx - \by \|, \quad \forall ~ \bx, \by \in \setX,
\]
and the corresponding constant $L_2$ is called a Lipschitz constant of $\nabla f$ on $\setX$.
Also, $f$ is simply said to have Lipschitz continuous gradient if $f$ has Lipschitz continuous gradient on the space on which $f$ is defined, i.e., either $\Rbb^n$ or $\Cbb^n$.
The Lipschitz continuity and Lipschitz continuous gradient conditions are automatically satisfied if $f$ is smooth and $\setX$ is compact, which is the case of our problem to be shown later.
Let us be precise here.
\begin{Fact} \label{fac:Lip_simple}
	Suppose that $\setX$ is compact.
	If $f$ is continuously differentiable, then $f$ is Lipschitz continuous on $\setX$.
	If $f$ is twice continuously differentiable, then $f$ has Lipschitz continuous gradient on $\setX$.
\end{Fact}
%\cite{Bertsekas1976}.

\section{Background}
\label{sec:background}

Consider the multiuser downlink scenario depicted in Fig.~\ref{fig:model}.
A base station (BS) equipped with a massive number of antennas is tasked with transmitting  symbol streams to a multitude of single-antenna users.
The BS deploys either one-bit DACs or phase shifters for low-cost and power-efficient implementations.
Assuming frequency-flat channels and the transmission time duration not exceeding the channel coherence time,
we can model the signals at the complex baseband level as
\begin{equation} \label{eq:rx_model}
y_{i,t} = \bh_i^T \bm \xi_t + \eta_{i,t}, \quad i=1,\ldots,K, t=1,\ldots,T,
\end{equation}
where
$y_{i,t}  \in \Cbb$ is the signal received by user $i$ at symbol time $t$;
$\bm\xi_t \in \Cbb^N$ is the multi-antenna signal transmitted by the BS at symbol time $t$;
$\bh_i \in \Cbb^N$ is the channel from the BS to user $i$;
$\eta_{i,t}$ is noise which is assumed to be circular complex Gaussian with mean zero and variance $\sigma_{\eta}^2$;
$T$ is the transmission block length;
$N$ is the number of transmit antennas;
$K$ is the number of users.
We express the transmitted signals as
\begin{equation} \label{eq:xt_model}
\bm \xi_t = \sqrt{\tfrac{P}{N}} \bu_t,
\quad
\bu_t \in \setU^N,
\end{equation}
where $P$ is the total transmission power;
$\bu_t$ is the normalized transmitted signal;
$\setU$ will be specified.
If the BS deploys one-bit DACs, we may choose
\begin{equation} \label{eq:ob_set}
\setU = \setU_{\textsf{1-bit}} \triangleq \Big\{ u_R + \jj u_I \Big| u_R, u_I \
\in \left\{ \pm \tfrac{1}{\sqrt{2}} \right\} \Big\}.
\end{equation}
Specifically, in the one-bit case, the real and imaginary parts of the transmitted signals are generated by one-bit DACs, and \eqref{eq:ob_set} characterizes that.
If the BS deploys phase shifters, or CE transmission, we may choose
\begin{equation} \label{eq:ce_set}
\setU =  \setU_{\textsf{CE}} \triangleq \{ u \in \Cbb \mid |u|=1 \}.
\end{equation}
The above characterization assumes that the phase shifters can generate a continuum of phase values over the whole phase range;
or, the phase shifters have fine phase resolutions which make \eqref{eq:ce_set} an accurate approximation.
If this is not the case, we can consider a discrete CE (DCE) model
\begin{equation} \label{eq:DCE_set}
\setU = \setU_{\textsf{DCE}} \triangleq \{ u = e^{\jj \left( \frac{2\pi}{M} m +  \frac{\pi}{M} \right) } \mid m=0,1,\ldots,M-1 \},
\end{equation}
for some even positive integer $M \geq 4$.
Eq.~\eqref{eq:DCE_set} assumes that the phase shifters admit uniform phase values.

\ifconfver
	\begin{figure*}
		\centering
		\includegraphics[width=.95\textwidth]{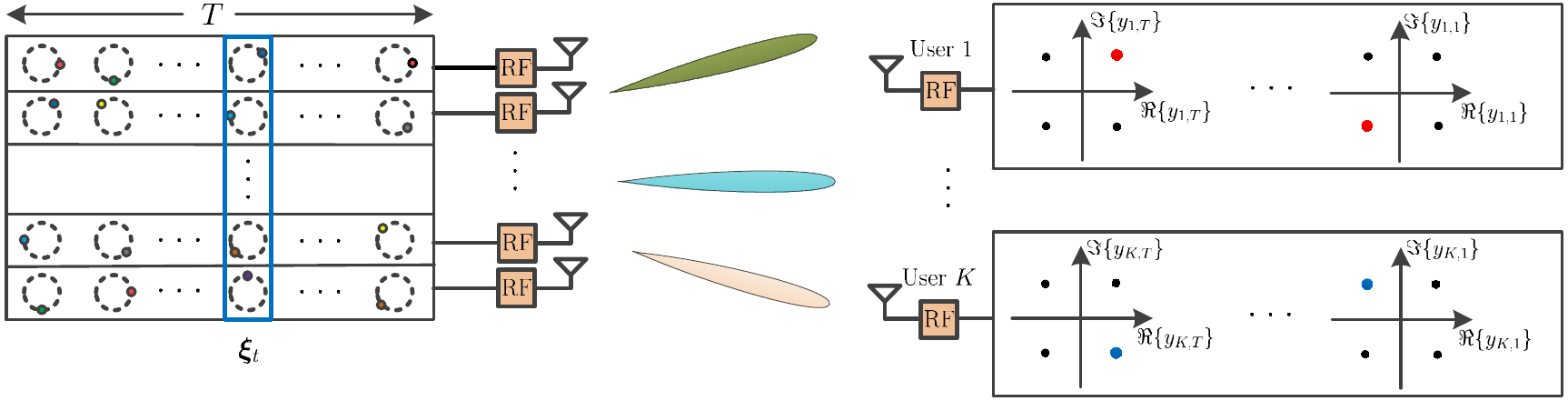}
		\caption{The scenario. }\label{fig:model}
	\end{figure*}
\else
	\begin{figure}
		\centering
		\includegraphics[width=.99\textwidth]{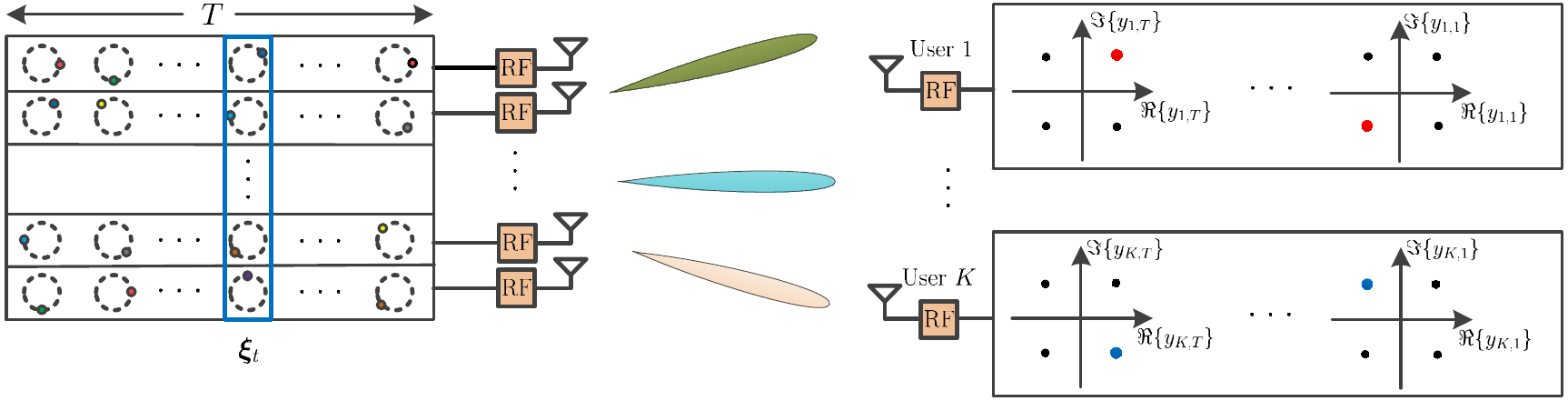}
		\caption{The scenario. }\label{fig:model}
	\end{figure}
\fi

We focus on the precoding problem.
Assume knowledge of the channels $\bh_i$'s at the BS.
The problem is to design $\bu_1, \ldots, \bu_T$ such that each user will see its designated symbol stream on $\{ y_{i,t} \}$.
To be specific, let $\{ s_{i,t} \}_{t=1}^T$ denote the symbol stream for user $i$.
We wish to have
\begin{equation} \label{eq:sigmod_temp}
\bh_i^T \bm\xi_t \approx s_{i,t}, \quad \forall ~ i,t.
\end{equation}
%In words, we want to shape the designated symbol streams at the users' sides.
Note that \eqref{eq:sigmod_temp} is only an illustration of the design aim,
and it is not exactly what we will do; this will be considered later.
We assume that the symbols are drawn from the QAM constellation, viz.,
\[
s_{i,t} \in \setS \triangleq \{ s_R  + \jj s_I \mid s_R, s_I \in \{ \pm 1, \pm 3, \ldots, \pm (2B-1) \} \},
\]
for some positive integer $B$ ($4B^2$ is the QAM size).
It should be emphasized that this is a nonlinear precoding problem wherein $\bm\xi_t$ is not necessarily a linear combination of the symbols $s_{i,t}$'s, or an outcome of linear precoding.
%{\blue Few works consider precoding designs within the whole transmission block \cite{Jacobsson2016}. By contrast, our design will operate on each transmission block rather than a single time slot at a time, which is appreciated in practice. }

%---------------------------------------------------------------------------
\section{Minimum SEP Problem Formulation}
\label{sec:sep}
Following the problem specified in the last section, we aim to formulate a precoding design that would minimize the impact of having unsuccessfully delivered symbol streams.
To this end, we choose the symbol-error probability (SEP) as our performance metric and establish a precoding design formulation.
The details are as follows.

%--------------------------
\subsection{SEP Characterization}

\begin{figure}
  \centering
  \includegraphics[width=0.6\linewidth]{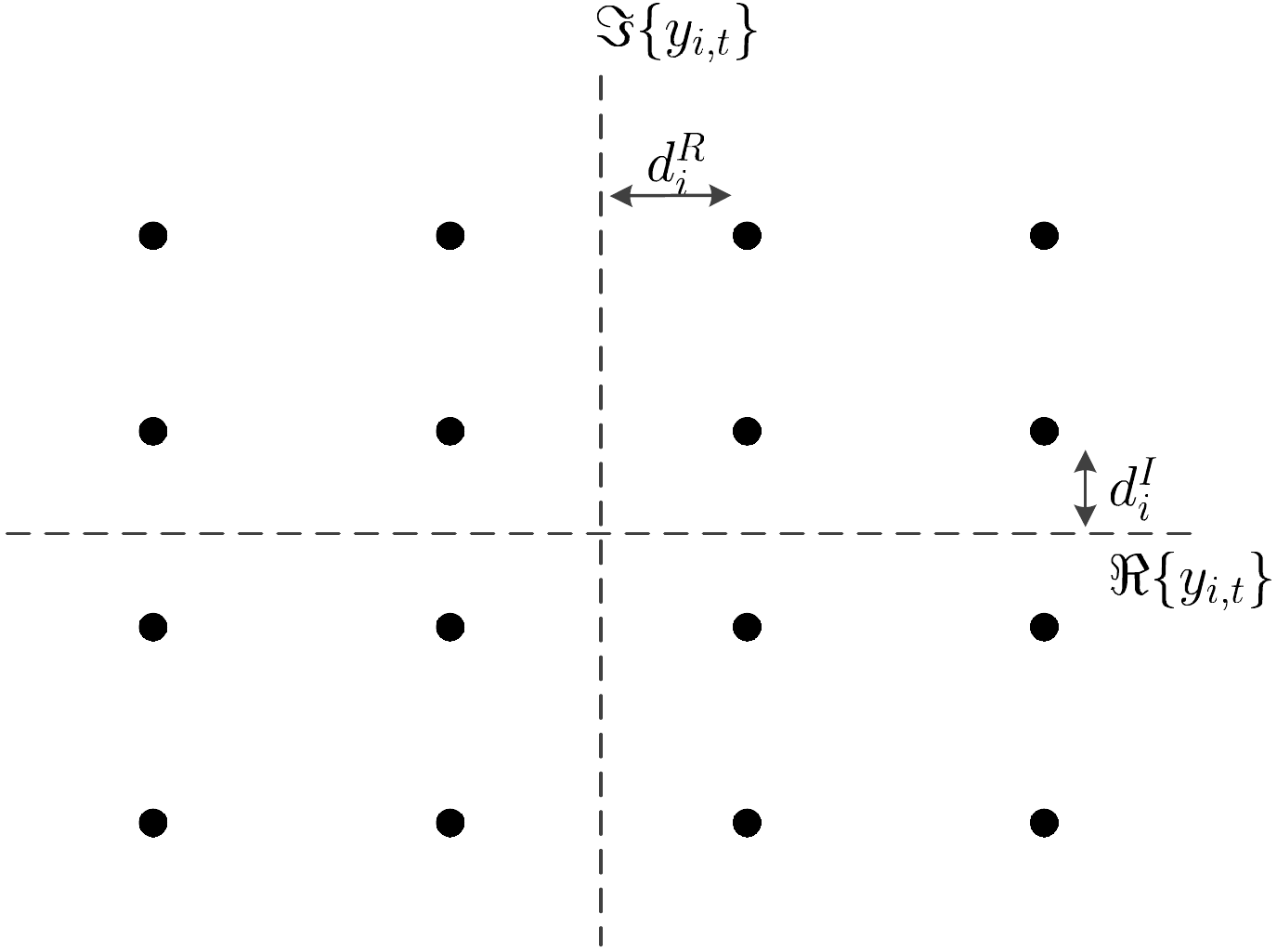}
  \caption{Illustration of $\dRi$ and $\dIi$ for $16$-QAM.}\label{fig:d}
\end{figure}

To work on SEP, it is first necessary to specify the symbol detection rule at the user side.
We assume that the users expect to receive
\[
y_{i,t} = \dRi \Re( s_{i,t} ) + \jj \cdot \dIi \Im( s_{i,t} ) + \eta_{i,t},
\]
where $\dRi$ and $\dIi$ describe the half inter-point spacings of the real and imaginary parts of the received QAM symbols at user $i$; see Fig \ref{fig:d}.
These inter-point spacings are determined by the BS, and the users are informed of their values during the training phase.
The users then detect the symbols via
\begin{equation} \label{eq:hat_s}
\hat{s}_{i,t} = \dec( \Re( y_{i,t} )/ \dRi ) + \jj \cdot \dec( \Im( y_{i,t} )/ \dIi ),
\end{equation}
where $\dec$ denotes the decision function for the set $\{ \pm 1, \pm 3, \ldots, \pm (2B-1) \}$.
Define
\begin{equation} \label{eq:SEP_def}
\begin{aligned}
\SEPR & = {\rm Pr}( \Re(\hat{s}_{i,t}) \neq \Re(s_{i,t}) \mid s_{i,t} ), \\
\SEPI & = {\rm Pr}( \Im(\hat{s}_{i,t}) \neq \Im(s_{i,t}) \mid s_{i,t} ),
\end{aligned}
\end{equation}
i.e., the error probabilities of real and imaginary parts of the symbol, conditioned on $s_{i,t}$.
We will simply call \eqref{eq:SEP_def} the SEPs although they are actually conditional.
It should be noted that
\ifconfver
\begin{align*}
 \max\{ \SEPR, \SEPI \} \leq & {\rm Pr}( \hat{s}_{i,t} \neq s_{i,t} | s_{i,t} ) \\
 \leq & 2 \max\{ \SEPR, \SEPI \},
\end{align*}
\else
\[
\max\{ \SEPR, \SEPI \} \leq {\rm Pr}( \hat{s}_{i,t} \neq s_{i,t} | s_{i,t} ) \leq 2 \max\{ \SEPR, \SEPI \},
\]
\fi
i.e., the  symbol-error probabilities can be effectively controlled by controlling the error probabilities of the real and imaginary parts of the symbol.
It can be shown from \eqref{eq:rx_model}, \eqref{eq:xt_model} and \eqref{eq:hat_s} that
\begin{equation} \label{eq:SEP_exact}
\arraycolsep=1.4pt\def\arraystretch{2.2}
\SEPR =
\left\{
\begin{array}{ll}
Q\left( \frac{ \sqrt{2} \bRi}{ \sigma_\eta } \right) + Q\left( \frac{ \sqrt{2} \cRi}{ \sigma_\eta } \right),
	& | \Re(s_{i,t}) | < 2B-1, \\
Q\left( \frac{ \sqrt{2} \cRi}{ \sigma_\eta } \right), & 	\Re(s_{i,t}) = 2B-1, \\
Q\left( \frac{ \sqrt{2} \bRi}{ \sigma_\eta } \right), & \Re(s_{i,t}) = -2B+1,
\end{array}
\right.
\end{equation}
where $Q(x) = \int_{x}^{\infty} \frac{1}{\sqrt{2\pi}} e^{-z^2/2} dz$,
\begin{equation} \label{eq:bc}
\begin{aligned}
\bRi & = \dRi - \left( \sqrt{\tfrac{P}{N}} \Re( \bh_i^T \bu_t ) - \dRi \Re(s_{i,t}) \right), \\
\cRi & = \dRi + \left( \sqrt{\tfrac{P}{N}} \Re( \bh_i^T \bu_t ) - \dRi \Re(s_{i,t}) \right).
\end{aligned}
\end{equation}
Also, the above result holds for $\SEPI$ if we replace ``$R$'' and ``$\Re$'' by ``$I$'' and ``$\Im$'', resp.
We shall skip the proof of \eqref{eq:SEP_exact}--\eqref{eq:bc} as it is almost a routine exercise on error probability analyses in digital communications \cite{goldsmith_2005}.

%--------------------------
\subsection{A Minimum SEP Formulation}

Our design is to provide uniformly good SEP performance over all users, specifically,
%This leads us to formulate the problem as
\begin{equation} \label{eq:P0}
\min_{\substack{ \bU \in \setU^{N \times T}, \\ \bd \geq \bzero }} ~
\max_{\substack{ i=1,\ldots,K, \\ t= 1,\ldots, T}} ~ \max\{ \SEPR, \SEPI \},
\end{equation}
where
$\bU = [~ \bu_1,\ldots,\bu_T ~]$, $\bd^R= [~ d_{1}^R,\ldots,d_{K}^R ~]^T$,
$\bd^I= [~ d_{1}^I,\ldots,d_{K}^I ~]^T$, $\bd = [~ (\bd^R)^T, ~ ({\bd^I})^T ~]^T$,
$\setU$ is given by \eqref{eq:ob_set} for one-bit precoding,
by \eqref{eq:ce_set} for CE precoding,
and by \eqref{eq:DCE_set} for DCE precoding.
As can be seen in \eqref{eq:P0}, we intend to achieve the aforementioned aim by minimizing the worst SEP among all the users and at all symbol times.
We should emphasize that Problem~\eqref{eq:P0} not only designs the precoder, it also optimizes the QAM inter-point spacings for best performance.

Problem~\eqref{eq:P0} has a drawback.
The functions $\SEPR, \SEPI$ in \eqref{eq:SEP_exact} do not admit simple expressions; they depend on the $Q$ function which has no closed form.
Instead of handling Problem~\eqref{eq:P0} directly, we choose to work on a closely related problem.
Consider the following fact which is easily shown from \eqref{eq:SEP_exact}.
\begin{Fact} \label{fact:SEP}
It holds that
\begin{equation} \label{eq:fact:SEP}
\begin{aligned}
Q\left( \frac{ \sqrt{2} \aRi }{ \sigma_\eta  } \right) & \leq \SEPR  \leq 2 Q\left( \frac{ \sqrt{2} \aRi }{ \sigma_\eta  } \right), \\
Q\left( \frac{ \sqrt{2} \aIi }{ \sigma_\eta  } \right) & \leq \SEPI  \leq 2 Q\left( \frac{ \sqrt{2} \aIi }{ \sigma_\eta  } \right),
\end{aligned}
\end{equation}
where
\[
\aRi = \min\{ \bRi, \cRi \}, \quad
\aIi = \min\{ \bIi, \cIi \}.
\]
Here, $\bRi$ and $\cRi$ are defined in \eqref{eq:bc};
$\bIi$ and $\cIi$ are defined by the same way as \eqref{eq:bc},
with ``$R$'' and ``$\Re$'' replaced by ``$I$'' and ``$\Im$'', resp.
\end{Fact}
Fact~\ref{fact:SEP} suggests that
we can suppress the SEPs by maximizing $\aRi$ and $\aIi$;
in fact, one may see that $\aRi/\sigma_\eta$ and $\aIi/\sigma_\eta$ appear like SNR terms in \eqref{eq:fact:SEP}.
In view of this,
we turn to
\begin{align}
& \min_{\substack{ \bU \in \setU^{N \times T}, \\ \bd \geq \bzero }} ~
\max_{\substack{ i=1,\ldots,K, \\ t= 1,\ldots, T}} ~ \max\{ -\aRi, -\aIi \} \nonumber  \\
 = & \min_{\substack{ \bU \in \setU^{N \times T}, \\ \bd \geq \bzero }} ~
\max_{\substack{ i=1,\ldots,K, \\ t= 1,\ldots, T}} ~ \max\{ -\bRi, -\cRi, -\bIi, -\cIi \}.
\label{eq:P1}
\end{align}
Now, the objective function is piece-wise linear and is much simpler than that of Problem~\eqref{eq:P0}.
Also, by applying \eqref{eq:fact:SEP} in Fact~\ref{fact:SEP} to Problem~\eqref{eq:P0}, and using the monotonicity of $Q$,
one can see that
Problem~\eqref{eq:P1} is both upper-bound and lower-bound approximations of Problem~\eqref{eq:P0}.
We also show the following result.
\begin{Prop} \label{fact:d_bnd}
	There exists an optimal solution $(\bU^\star,\bd^\star)$ to Problem~\eqref{eq:P1} such that $\bd^\star \leq \bm\rho$, where $\rho_{i}= \rho_{i+N} = \sqrt{P/N} \| \bh_i \|_1$ for $i=1,\ldots,N$.
\end{Prop}
The proof of Proposition~\ref{fact:d_bnd} is relegated to Appendix~\ref{appen:proof_d_bnd}.

There is still an issue, though a lesser one.
The objective function in \eqref{eq:P1} is non-smooth.
There are various ways to tackle non-smooth optimization problems,
and we resort to smooth approximation which has the advantage of allowing us to access powerful tools in smooth optimization.
Specifically, we apply the log-sum-exponential (LSE) approximation.
Let ${\rm lse}(\bx) = \sigma \log( \sum_{i=1}^n e^{x/\sigma} )$ where $\sigma > 0$, $\bx \in \Rbb^n$.
It is known that ${\rm lse}(\bx)$ approximates $\max\{x_1,\ldots,x_n\}$ with an accuracy that improves as $\sigma$ decreases, and the approximation is tight as $\sigma \rightarrow 0$ \cite{CVX}.
By applying the LSE approximation to Problem~\eqref{eq:P1}, we arrive at our final formulation as follows.

\medskip\medskip
\noindent
\boxed{
\begin{minipage}{\linewidth}
Given a smoothing parameter $\sigma > 0$, solve
\begin{equation} \label{eq:P_main}
\min_{\substack{ \bU \in \setU^{N \times T}, \\ \bzero \leq \bd \leq \bm\rho }} ~
f(\bU,\bd) \triangleq \sigma \log \left( \sum_{t=1}^T \sum_{i=1}^K  f_{i,t}(\bu_t,\bd) \right),
\end{equation}
where
$\bm\rho$ is defined in Proposition~\ref{fact:d_bnd};
\begin{equation} \label{eq:f_def}
f_{i,t}(\bu_t,\bd) = e^{-\frac{\bRi}{\sigma}} + e^{-\frac{\bIi}{\sigma}} + e^{-\frac{\cRi}{\sigma}} + e^{-\frac{\cIi}{\sigma}};
\end{equation}
and recall that $\bRi$ and $\cRi$ are defined in \eqref{eq:bc}, and $\bIi$ and $\cIi$ are similarly defined for ``$I$''.
\end{minipage}
}

\medskip\medskip
We also have the following remark.

\begin{Remark}
%While we will only concentrate on the minimum worst SEP design,
%It is worthwhile to mention that
Problem~\eqref{eq:P_main} can also be interpreted as a design that attempts to minimize the average SEP.
Let $\overline{\sf SEP} = \frac{1}{2KT}\sum_{t=1}^T \sum_{i=1}^K (\SEPR + \SEPI)$ be the average SEP.
By applying the inequality $Q(x) \leq 0.5 e^{-\sqrt{2/\pi} x}$
 to \eqref{eq:SEP_exact},\footnote{It is known that $Q(x) \leq 0.5 e^{-\sqrt{2/\pi} x}$ for $x \geq 0$ \cite{verdu1998multiuser}. For the case of $x \leq 0$, we prove it as follows. Let $f(x) = 0.5 e^{-\sqrt{2/\pi} x} - Q(x)$.
 By examining the derivative of $f$, one can verify that $f(x)$ is decreasing on $x \leq 0$. Since $f(0) = 0$, we have $f(x) \geq 0$ for $x \leq 0$. This implies $0.5 e^{-\sqrt{2/\pi} x} \geq Q(x)$ for $x\leq 0$.}
we see that
\[
\overline{\sf SEP} \leq \frac{1}{4KT} \sum_{t=1}^T \sum_{i=1}^K  f_{i,t}(\bu_t,\bd),
\]
for $\sigma= (\sqrt{\pi}/2) \sigma_\eta$.
This implies that Problem~\eqref{eq:P_main} tends to suppress the average SEP if we choose $\sigma= (\sqrt{\pi}/2) \sigma_\eta$.
\end{Remark}

%--------------------------
%--------------------------

\section{A Negative Square Penalty Method}
\label{sec:NSP}

Our next problem is to build an algorithm for the precoding design  formulated  in Problem~\eqref{eq:P_main}.
Finding a working algorithm for \eqref{eq:P_main} is not trivial.
Problem~\eqref{eq:P_main} has a convex smooth objective function, but it  has a non-convex constraint set;
the set $\setU$ is a manifold for the CE case and is discrete for the one-bit and DCE cases.
Dealing with such constraints is known to be difficult.
In this section we will first develop a method that will allow us to transform the precoding problem to a convex constrained problem with a non-convex smooth objective function.
Then, in the next section, we will custom-build a fast algorithm for the transformed problem.

\begin{figure}
	\centering
	\begin{subfigure}[b]{0.3\linewidth}
		\includegraphics[width=\textwidth]{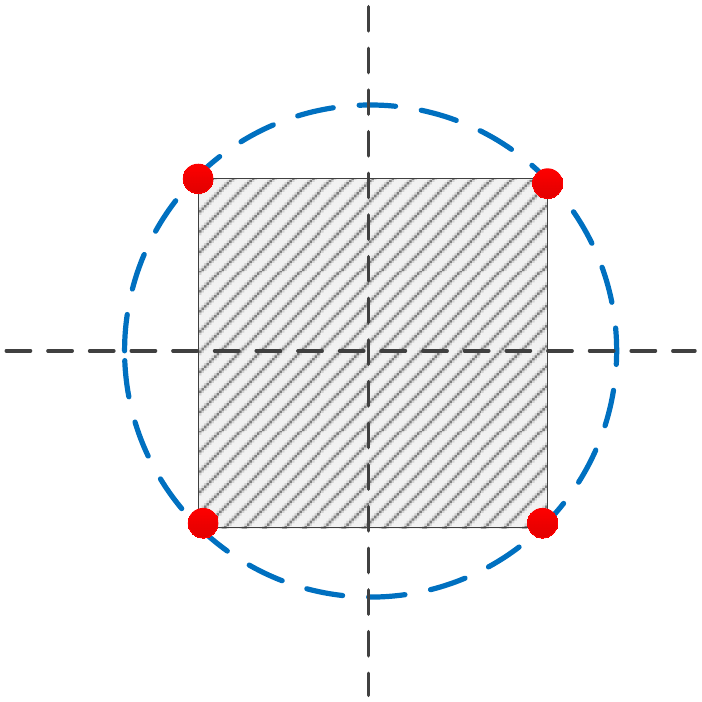}
		\caption{One-bit set}
		\label{fig:obset}
	\end{subfigure}
	~ %add desired spacing between images, e. g. ~, \quad, \qquad, \hfill etc.
	\begin{subfigure}[b]{0.3\linewidth}
		\includegraphics[width=\textwidth]{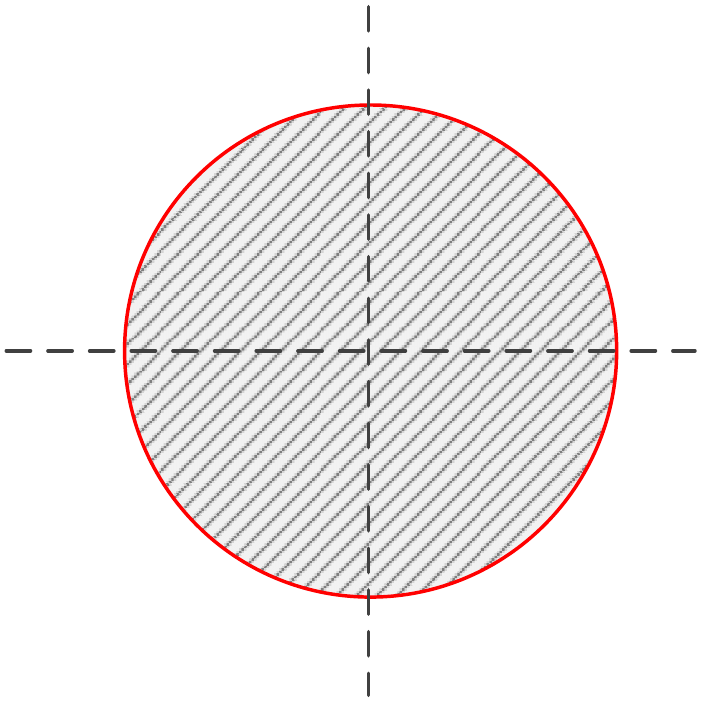}
		\caption{CE set}
		\label{fig:CEset}
	\end{subfigure}
	~ %add desired spacing between images, e. g. ~, \quad, \qquad, \hfill etc.
	\begin{subfigure}[b]{0.3\linewidth}
		\includegraphics[width=\textwidth]{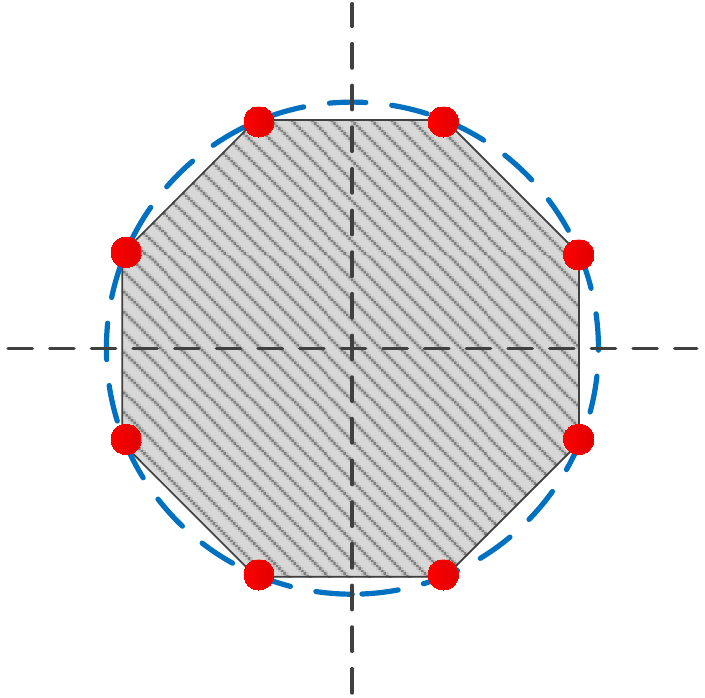}
		\caption{DCE set; $M=8$}
		\label{fig:QCEset}
	\end{subfigure}
	\caption{Illustration of $\setU$ and $\setbarU$. Red: $\setU$, shaded area: $\setbarU$.}\label{fig:convU}
\end{figure}

The method to be proposed considers optimization problems that take the form
\begin{equation} \label{eq:NEP_orig}
\min_{\bu \in \setU^n} ~ f(\bu),
\end{equation}
where $f: \Cbb^n \rightarrow \Rbb$ is the objective function; $\setU$ is either the one-bit set in \eqref{eq:ob_set}, the CE set in \eqref{eq:ce_set}, or the DCE set in \eqref{eq:DCE_set}.
As mentioned, the constraint set $\setU^n$ is hard to deal with.
The proposed method hinges on the use of a negative square penalty (NSP),
specifically,
\begin{equation} \label{eq:NEP}
\min_{\bu \in \setbarU^n} ~ F_\lambda(\bu) \triangleq f(\bu) - \lambda \| \bu \|^2,
\end{equation}
where $\setbarU = \conv  \setU$ is the convex hull of $\setU$; $\lambda > 0$ is a penalty parameter.
The idea is simple: From the illustration in Fig.~\ref{fig:convU},
one can see that the set of all extreme points of $\setbarU$ is $\setU$ itself.
The penalty term $-\lambda \| \bu \|^2$ is used to push each $u_i$ to an extreme point of $\setbarU$.
It is worthwhile to note that Problem~\eqref{eq:NEP} has a convex constraint set,
though we should also point out that $F_\lambda$ is generally non-convex even if $f$ is convex.
We will see in the next section that
we can custom-build very efficient first-order methods to handle
Problem~\eqref{eq:NEP} if $f$ is smooth.

It is natural to question whether the NSP problem~\eqref{eq:NEP} can be an exact reformulation of the original problem~\eqref{eq:NEP_orig}.
This is answered in the following theorem.
\begin{Theorem} \label{thm:NEP}
	Suppose that $f$ is Lipschitz continuous on $\setbarU^n$
	(see Section~\ref{sec:notations} for the definition of Lipschitz continuity).
	Then there exists a constant $\bar{\lambda} > 0$ such that for any $\lambda > \bar{\lambda}$,
	any (globally) optimal solution to Problem~\eqref{eq:NEP} is also a (globally) optimal solution to Problem~\eqref{eq:NEP_orig};
	the converse is also true.
	Specifically, we have $\bar{\lambda} = \sqrt{2}L$ for the one-bit case, $\bar{\lambda} = L$ for the CE case, and $\bar{\lambda} = L/\sin(\pi/M)$ for the DCE case, where $L$ is a Lipschitz constant of $f$ on $\setbarU^n$.
\end{Theorem}
The proof of Theorem~\ref{thm:NEP} is shown in Appendix~\ref{appen:proof_thm_NEP}.
%Note that as discussed in Section \ref{sec:notations}, the assumption of $f$ being Lipschitz continuous on $\setbarU^n$---which is compact---is considered mild.
%; see  Fact~\ref{fac:Lip_simple}.
Theorem~\ref{thm:NEP} reveals that for a sufficiently large $\lambda > 0$, the optimal solution sets of Problems~\eqref{eq:NEP_orig} and \eqref{eq:NEP} are equivalent.
Note that this equivalence does not require fine tuning of $\lambda$, and a sufficiently large $\lambda$ suffices.
One may further question whether a similar relationship can be shown for locally optimal solutions.
%Fact~\ref{fact:NEP_counterexample} considers a non-smooth instance.
%For smooth $f$, the following result can be shown.
Our answer is as follows.
\begin{Theorem}\label{theorem:NEP_smooth}
	Suppose that $f$ has Lipschitz continuous gradient on $\setbarU^n$
	 (see Section~\ref{sec:notations} for the definition of Lipschitz continuous gradient).
	Then there exists a constant $\bar{\lambda}$ such that for any $\lambda > \bar{\lambda}$, any locally optimal solution to Problem~\eqref{eq:NEP} must be a feasible solution  to Problem~\eqref{eq:NEP_orig}. Specifically, we have $\bar{\lambda}=L/2$ where $L$ is a Lipschitz constant of $\nabla f$ on $\bar{\setU}^n$.
%{\blue Moreover, for CE case, any locally optimal solution to Problem~\eqref{eq:NEP} must be locally optimal to Problem~\eqref{eq:NEP_orig}.}
	%any globally optimal solution to Problem~\eqref{eq:NEP} is also a globally optimal solution to Problem~\eqref{eq:NEP_orig}. Moreover,
\end{Theorem}
The proof of Theorem~\ref{theorem:NEP_smooth} is shown in Appendix~\ref{app:proof_NEP_eqv_smooth}.
Theorem~\ref{theorem:NEP_smooth} provides the implication that for smooth $f$ and for a sufficiently large $\lambda$, any stationary point of the NSP problem~\eqref{eq:NEP} that does not lie in $\setU^n$ would either be a saddle point or a local maximum.
Intuitively one may argue that for algorithms such as descent-based methods, converging to such points is not too likely.

We will apply the NSP method to the precoding problem in the next section.
Here we discuss a few more fundamental aspects with NSP, and readers who are more interested in the precoding application may jump to the next section.

\begin{Remark}
Upon a closer look at Theorem~\ref{thm:NEP}, readers may find that the bound $\bar{\lambda}$ in the DCE case does not look consistent with that in the CE case.
In particular, intuitively one may expect that the result  $\bar{\lambda} = L/\sin(\pi/M)$ in the DCE case should converge to the result $\bar{\lambda} = L$ in the CE case as $M \rightarrow \infty$, and yet this is not the situation.
Readers who have examined the proof would realize that when $M$ is large, it takes a stronger penalty to push the solution from the face of $\setbarU^n$ to an extreme point.
On the other hand, we can prove the following result.
\begin{Corollary} \label{Cor:NEP}
	Consider the DCE case under the same settings as in Theorem~\ref{thm:NEP}.
	Let $\hat{\bu}$ be an optimal solution to Problem~\eqref{eq:NEP},
	let $\tilde{\bu} = \Pi_{\setU^n}(\hat{\bu})$ be the DCE-rounded point of $\hat{\bu}$,
	and let $f^\star = \min_{\bu \in \setU^n} f(\bu)$.
	For $\lambda \geq L/\cos(\pi/M)$, with $L$ a Lipschitz constant of $f$ on $\setbarU^n$, we have
	\[
	f^\star \leq f(\tilde{\bu}) \leq f^\star + L \sqrt{n} \sin(\pi/M).
	\]
	In particular, if $M \rightarrow \infty$, we have $f(\tilde{\bu}) \rightarrow f^\star$ and $\cos(\pi/M) \rightarrow 1$.
\end{Corollary}
Corollary~\ref{Cor:NEP} suggests that the DCE-rounded point of an optimal solution to the NSP problem is a good approximate solution to the original problem when $M$ is large.
Corollary~\ref{Cor:NEP} is a consequence of Theorem~\ref{thm:NEP} and the proof is  shown in Appendix~\ref{appen:Cor:NEP}.
\end{Remark}

\begin{Remark}
The local optimality result in Theorem~\ref{theorem:NEP_smooth} is based on the premise that $f$ must at least be continuously differentiable.
One may wonder if the same result holds for non-smooth $f$.
Unfortunately, a counter-example can be found.
\begin{Fact}\label{fact:NEP_counterexample}
	Consider Problem~\eqref{eq:NEP} with $n=1$, $f(u) = |u|$ and $\setU = \{ u \in \Cbb \mid |u|=1 \}$. It holds that for any finite $\lambda > 0$, $u=0$ is
	%There exists an instance of $f$, $\setU$ for which
	a locally optimal solution to Problem~\eqref{eq:NEP}, but $u=0$ is infeasible to Problem~\eqref{eq:NEP_orig}.
\end{Fact}
The proof of Fact~\ref{fact:NEP_counterexample} is  shown in Appendix \ref{appen:NEP_counterexample}.
\end{Remark}

\begin{Remark}
%While we will focus only on the NSP method,
%with an emphasis on tackling the precoding problem,
We should mention related methods.
In nonlinear programming, the exact penalty methods are well known \cite{nocedal2000}.
Taking the DCE case as an example,
one may apply the exact penalty method to reformulate Problem~\eqref{eq:NEP_orig} as
%an exact penalty formulation
\begin{equation}
\label{eq:exact_penalty_example}
\min_{\bu \in \Cbb^n} ~ f(\bu) + \lambda \sum_{i=1}^n \left| 1 - (u_i e^{-\jj \frac{\pi}{M}} )^M \right|.
\end{equation}
Note that $u_i \in \setU$ is equivalent to $ (u_i e^{-\jj \frac{\pi}{M}} )^M = 1$, and the aim of the above penalty is to enforce $ (u_i e^{-\jj \frac{\pi}{M}} )^M = 1$.
The upshot of the exact penalty problem \eqref{eq:exact_penalty_example} is that it is unconstrained,
but the downside is that the penalty function involves higher-order polynomials.
In comparison, the NSP problem~\eqref{eq:NEP} is constrained, but its penalty is quadratic regardless of $M$.
Moreover, we should recognize the penalty method in \cite{Yuan2016}.
This method is somehow similar to our NSP method, although it considered binary optimization problems only and did not consider the CE and DCE cases here.
Expert readers would find that the underlying ideas, the specific penalties and the subsequent optimality analyses (Theorems~\ref{thm:NEP}--\ref{theorem:NEP_smooth} for NSP) of the two works are different,
and that the NSP method seems more straightforward.
\end{Remark}

%----------------------------------
%----------------------------------
\section{Gradient Extrapolated Majorization-Minimization}
\label{sec:gemm}

We now come to the final part of our development, namely, custom-building an algorithm for the NSP-transformed formulation of the precoding problem \eqref{eq:P_main}.

\subsection{The Main Idea}
\label{sec:gemm_main_idea}

By applying the NSP method in Section~\ref{sec:NSP}, we equivalently reformulate the precoding problem \eqref{eq:P_main} as
\begin{equation} \label{eq:P_app}
\min_{ { \bU \in \setbarU^{N \times T},   \bd\in \setD }} ~
F_{\lambda}(\bU,\bd) \triangleq f(\bU,\bd) -\lambda \sum_{t=1}^{T} \| \bu_t \|^2,
\end{equation}
where $\setD\triangleq \{ \bd \mid \bzero \leq \bd \leq \bm\rho \}$, and $\lambda > 0$ is assumed to be sufficiently large.\footnote{As a technical note, the precoding problem \eqref{eq:P_main} does not take exactly the same form as the problem considered for NSP, i.e., Problem~\eqref{eq:NEP_orig}.
Specifically, \eqref{eq:P_main} has an extra decision variable $\bd$.
However, it can be easily shown that the same NSP concepts and optimality results (Theorems~\ref{thm:NEP}--\ref{theorem:NEP_smooth}) apply.}
%It should be noted that the precoding problem \eqref{eq:P_main} does not take exactly the same form as the problem considered for NSP (cf. Problem~\eqref{eq:NEP_orig}),
%where \eqref{eq:P_main} has an extra decision variable $\bd$,
% lying on a convex compact set $\setD$,
%but
%it is easy to show that the same NSP concepts and optimality results (Theorems~\ref{thm:NEP}--\ref{theorem:NEP_smooth}) apply.
For ease of exposition of the idea, let us notationally simplify Problem~\eqref{eq:P_app} by rewriting it as
\begin{equation} \label{eq:GEMM_main_1}
\min_{  \bx \in \setX } ~
F_{\lambda}(\bx) = f(\bx) -\lambda \| \bx_{1:T} \|^2,
\end{equation}
where $\bx=  (\bx_1,\ldots,\bx_T,\bx_{T+1})$, $\bx_i = \bu_i$ for $i=1,\ldots,T$, $\bx_{T+1} = \bd$, $\setX = \setbarU^N \times \cdots \times \setbarU^N \times \setD$,
and $\bx_{1:T}= (\bx_1,\ldots,\bx_T)$.
We tackle Problem~\eqref{eq:GEMM_main_1} by the majorization-minimization (MM) method.
The MM method, in its general form, is given by
\begin{eqnarray} \label{eq:MM_loop}
\bx^{k+1} = \arg \min_{\bx \in \setX} G_\lambda(\bx| \bx^k), \quad k=0,1,2,\ldots,
\end{eqnarray}
where $G_\lambda(\cdot | \bar{\bx})$ denotes a majorant of $F_\lambda$ at $\bar{\bx}$, i.e., it satisfies $G_\lambda(\bx|\bar{\bx}) \geq F_\lambda(\bx)$ for all $\bx, \bar{\bx} \in \setX$ and  $G_\lambda(\bar{\bx}|\bar{\bx}) = F_\lambda(\bar{\bx})$ for all $\bar{\bx} \in \setX$ \cite{hunter2004tutorial}.
It is easy to derive a majorant for our problem.
Since $\| \bx \|^2 \geq \| \bar{\bx} \|^2 + 2 \langle \bar{\bx}, \bx - \bar{\bx} \rangle$ for any $\bx, \bar{\bx}$, we have
\begin{equation} \label{eq:majorizer}
\begin{aligned}
F_\lambda(\bx) & \leq f(\bx) - 2 \lambda \langle \bar{\bx}_{1:T}, {\bx}_{1:T} - \bar{\bx}_{1:T} \rangle - \lambda \| \bar{\bx}_{1:T} \|^2   \\
& \triangleq G_\lambda(\bx|\bar{\bx});
\end{aligned}
\end{equation}
it is also obvious that $G_\lambda(\bar{\bx}|\bar{\bx}) = F_\lambda(\bar{\bx})$.
Note that the majorant \eqref{eq:majorizer} is smooth and convex, and $\nabla_\bx G_\lambda(\bar{\bx} | \bar{\bx}) = \nabla_\bx F_\lambda(\bar{\bx})$.

To perform MM, we also need to compute the optimal solutions in \eqref{eq:MM_loop}.
The problems in \eqref{eq:MM_loop} are convex smooth optimization problems, and we choose the Nesterov or FISTA-type accelerated projected gradient (APG) method to solve them.
The APG method, as well as its predecessor, projected gradient (PG), are suitable for problems in which the projection operation $\Pi_\setX$ is easy to compute.
Also, APG is known to converge much faster than PG if the problem is convex.
The core concepts and technical details of the PG and APG methods have been extensively covered in the literature \cite{beck2017first};
here we consider application and shall be concise.
The APG method for solving $\min_{\bx \in \setX} G_\lambda(\bx|\bar{\bx})$ is
\begin{align*}
\bx^{i+1} = \Pi_\setX \left( \bz^i - \frac{1}{\beta_i} \nabla_\bx G_\lambda(\bz^i| \bar{\bx}) \right), \quad i=0,1,2,\ldots
\end{align*}
where $1/\beta_i > 0$ represents the step size; $\bz^i$ is an extrapolated point and is given by
\begin{equation} \label{eq:z_exa}
\bz^i = \bx^i +\alpha_i (\bx^i -\bx^{i-1}),
\end{equation}
with	
\begin{equation} \label{eq:alpha_exa}
 \alpha_i=\frac{\xi_{i-1}-1}{\xi_{i}}, \quad \xi_i=\frac{1+\sqrt{1+4 \xi_{i-1}^2}}{2},
\end{equation}
and with $\xi_{-1}=0, \bx^{-1} = \bx^0$.
Note that $\{ \alpha_i \}_{i \geq 0}$ is called the extrapolation sequence, and that if we replace \eqref{eq:z_exa} by $\alpha_i = 0$ for all $i$, the method reduces to the PG method.
Also, the step size $1/\beta_i$ is chosen such that $\bx^{i+1}$ satisfies the so-called descent property
\ifconfver
	\begin{align}
	G_\lambda(\bx^{i+1}|\bar{\bx}) & \leq G_\lambda(\bz^i|\bar{\bx})
		+ \langle \nabla_\bx G_\lambda(\bz^i| \bar{\bx}), \bx^{i+1}- \bz^i \rangle  \nonumber \\
		&	\quad	+ \frac{\beta_i}{2} \| \bx^{i+1}- \bz^i \|^2. \label{eq:desc_exa}
	\end{align}
\else
	\begin{align}  \label{eq:desc_exa}
	G_\lambda(\bx^{i+1}|\bar{\bx}) \leq G_\lambda(\bz^i|\bar{\bx})
	+ \langle \nabla_\bx G_\lambda(\bz^i| \bar{\bx}), \bx^{i+1}- \bz^i \rangle
	+ \frac{\beta_i}{2} \| \bx^{i+1}- \bz^i \|^2.
	\end{align}
\fi
We employ the backtracking line search method \cite{beck2017first} to compute such a $\beta_i$.

The above MM method is not exactly what we do.
The MM method in \eqref{eq:MM_loop} requires solving an optimization problem in an exact fashion at every iteration, and that is computationally expensive.
We consider an inexact MM where every MM iteration is a one-step APG update; specifically,
\begin{equation} \label{eq:GEMM_loop}
\bx^{k+1} =  \Pi_\setX \left( \bz^k - \frac{1}{\beta_k} \nabla_\bx G_\lambda(\bz^k| \bx^k) \right), \quad k=0,1,2,\ldots,
\end{equation}
where $\bz^k$ and $\beta_k$ are obtained by the same way as in \eqref{eq:z_exa}--\eqref{eq:desc_exa}; for the sake of clarity we have
\begin{equation} \label{eq:z_exa2}
\bz^k = \bx^k +\alpha_k (\bx^k -\bx^{k-1}),
\end{equation}
where $\{ \alpha_k \}_{k \geq 0}$ is the same sequence as in \eqref{eq:alpha_exa}, and $\beta_k$ is chosen such that
\ifconfver
	\begin{align}
	G_\lambda(\bx^{k+1}|\bx^k) & \leq G_\lambda(\bz^k|\bx^k)
	+ \langle \nabla_\bx G_\lambda(\bz^k|\bx^k), \bx^{k+1}- \bz^k \rangle \nonumber \\
	& \quad + \frac{\beta_k}{2} \| \bx^{k+1}- \bz^k \|^2. \label{eq:desc_exa2}
	\end{align}
\else
	\begin{align}  \label{eq:desc_exa2}
	G_\lambda(\bx^{k+1}|\bx^k) \leq G_\lambda(\bz^k|\bx^k)
	+ \langle \nabla_\bx G_\lambda(\bz^k|\bx^k), \bx^{k+1}- \bz^k \rangle
	+ \frac{\beta_k}{2} \| \bx^{k+1}- \bz^k \|^2.
	\end{align}
\fi
We name the method in \eqref{eq:GEMM_loop} {\em gradient extrapolated MM} (GEMM).
Our empirical studies suggest that GEMM is much faster than MM (implemented via APG) in terms of convergence speed, and GEMM appears to give satisfactory SEP performance most of the time.
This will be illustrated in the numerical simulation section in Section~\ref{sec:simulation}.

%----
\subsection{Convergence Guarantee of GEMM}

Our interest lies in the application of GEMM to precoding, and we will elaborate on the implementation details in the next subsection.
On the other hand, we can say about its convergence in the theoretical sense.
Consider a more general context where we deal with an optimization problem of the form
\[
\min_{\bx \in \setX} F(\bx),
\]
in which $F$ is differentiable, and $\setX$ is convex, non-empty and closed.
We apply the GEMM method  to this problem by replacing $G_\lambda$ in \eqref{eq:GEMM_loop}--\eqref{eq:desc_exa2} by some majorant $G$ of $F$,
and we question whether $\{ \bx^k \}_{k \geq 0}$ would possess any stationarity guarantees.
We should point out that GEMM does not exhibit the monotonic non-increasing property $F(\bx^0) \geq F(\bx^1) \geq F(\bx^2) \geq \cdots$, owing to  extrapolation.
Many first-order convergence analysis results assume some form of sufficient decrease of the objective function, and they are not applicable to GEMM.
In fact,  convergence analyses for non-convex first-order methods involving accelerated proximal gradient or APG are challenging, with open questions remaining; see the discussion in \cite{jin2017accelerated}.
Here, we take ideas from \cite{Yin2013}, which deals with block coordinate descent and not MM, to handle technical issues arising from extrapolation.
The following theorem describes the result.

\begin{Theorem} \label{thm:GEMM_conv}
	Consider the context described above.
	Suppose that
	\begin{enumerate}[1.]
		\item $F^\star \triangleq \inf_{\bx \in \setX} F(\bx)$ is finite;
		\item $F$ has Lipschitz continuous gradient with constant $L_F$
		 (see Section~\ref{sec:notations} for the definition of Lipschitz continuous gradient);
		\item $G$ satisfies i) $G(\bx|\bar{\bx}) \geq F(\bx)$ for any $\bx, \bar{\bx}$, ii) $G(\bar{\bx}|\bar{\bx}) = F(\bar{\bx})$ for any $\bar{\bx}$,
		that iii) $G(\cdot|\bar{\bx})$ is differentiable and has $\nabla_\bx G(\bar{\bx}|\bar{\bx}) = \nabla_\bx F(\bar{\bx})$ for every $\bar{\bx} \in \setX$, and that iv) $G(\cdot|\bar{\bx})$ has Lipschitz continuous gradient with constant $L_G$ for every $\bar{\bx} \in \setX$.
	\end{enumerate}
	Also, suppose that $\alpha_k$ and $\beta_k$ in GEMM in \eqref{eq:GEMM_loop}--\eqref{eq:desc_exa2} satisfy
		\[ 0\leq\alpha_k\leq \bar{\alpha} , \quad c_1 L_G \leq \beta_{k} \leq c_2 L_G,~~\forall~k \]
	for some constants $\bar{\alpha} = \sqrt{{c_1(1-\mu)}/{c_2}}$ with $0<\mu \leq 1$, $0< c_1<1$ and $1<c_2 <\infty$.
	Then, GEMM is guaranteed to find an $\epsilon$-stationary point in $\mathcal{O}(1/\epsilon^2)$ iterations.
	Specifically, it holds that
	\begin{equation*}
	\min_{k'=1,\ldots,k+1}{\rm dist} ( \bzero,\nabla F(\bx^{k'})+ \partial \ind_{\setX}(\bx^{k'}) ) \leq \frac{C}{\sqrt{k}},
	\end{equation*}
	where $C= C_1 \sqrt{ 8 (F(\bx^{0})- F^{\star})/ (c_1 L_G \mu)}$,
	$C_1 = \max\{ \bar{\alpha}(1+c_2)L_G, L_F+c_2L_G\}$.
%	$C= \sqrt{8 (\max\{ \bar{\alpha}(1+c_2)L_G, L_F+c_2L_G\})^2 (F(\bx^{0})- F^{\star})/ (c_1 L_G \mu)}$.
\end{Theorem}
The proof of Theorem~\ref{thm:GEMM_conv} is  shown in Appendix~\ref{app:proof_convergence}.
%We should mention that our proof takes ideas from \cite{Yin2013} to handle technical issues arising from extrapolation.
A key difference of our convergence analysis is that we prove convergence rate, rather than asymptotic convergence as in the previous work \cite{Yin2013}.
%Also, the convergence analysis of GEMM seems to have not been considered in prior work.

%\begin{Remark}
%	We should discuss works related to GEMM.
%	While we should first emphasize that the development of GEMM, along side with NSP, is new for one-bit/CE/DCE precoding,
%	the notion of employing inexact updates as an attempt to reduce complexity is not completely new from the perspective of optimization in signal processing and machine learning.
%	In the optimization context, there has been much interest in proving conditions under which an inexact update scheme can guarantee convergence to a stationary point.
%	For example,
%	\cite{hong2017iteration}  studied a block coordinate descent (BCD) framework under inexact updates by MM and proximal gradient,
%	while \cite{Yin2013} also considered inexact BCD, but under accelerated proximal gradient; note that convergence analyses for schemes involving accelerated proximal gradient or APG are challenging for non-convex problems, with open questions remaining (see the discussion in \cite{jin2017accelerated}).
%	Curiously, it seems that convergence analyses of GEMM have not been dealt with in the prior work.
%	Our proof of Theorem~\ref{thm:GEMM_conv} takes ideas from \cite{Yin2013} to handle technical issues arising from the use of APG.
%	A key difference of our convergence analysis is that we prove convergence rate, rather than asymptotic convergence as in the previous work.
%\end{Remark}

\subsection{Implementation Details}

Let us complete our work by filling in the implementation details.
Following the GEMM method introduced in Section~\ref{sec:gemm_main_idea}, we obtain the GEMM algorithm for one-bit/CE/DCE precoding in Algorithm~\ref{Al:GEMM_DCE}.
There are two key operations that require further explanation.
The first is the projection operations.
To facilitate our description, let $[ \bx ]_\ba^\bb$ define the element-wise thresholding operator; i.e., $\by = [ \bx ]_\ba^\bb \Longleftrightarrow y_i = \min\{ b_i, \max\{x_i,a_i\} \}$ for all $i$.
The projection $\Pi_\setD$ is simply
\[
\Pi_\setD(\bd) = [ \bd ]_\bzero^{\bm\rho}.
\]
The projection $\Pi_{\setbarU^{N \times T}}$ is merely the element-wise projection onto $\setbarU$, and thus it suffices to consider $\Pi_{\setbarU}$.
It is easy to see that for the one-bit case,
\[
\Pi_{\setbarU_{\sf 1-bit}}(u)  = \left[ \Re(u) \right]_{-1/\sqrt{2}}^{1/\sqrt{2}} + \jj \left[ \Im(u) \right]_{-1/\sqrt{2}}^{1/\sqrt{2}},
\]
and that for the CE case,
\[
\Pi_{\setbarU_{\sf CE}}(u)  =  \left\{
\begin{array}{ll}
u, & |u| \leq 1, \\
u/|u|, & |u| > 1.
\end{array}
\right.
\]
The projection for the DCE case is less obvious.
At first glance, one would be tempted to solve the projection by rewriting the constraint $u \in \setbarU_{\sf DCE}$ as linear inequalities (see, e.g., \cite{Jedda2018}) and then by calling a convex optimization solver to find the solution to $\Pi_{\setbarU_{\sf DCE}}(u)$.
As it turns out, it can be shown that $\Pi_{\setbarU_{\sf DCE}}(u)$ admits the closed-form expression
\begin{equation} \label{eq:DCE_round}
\Pi_{\setbarU_{\sf DCE}}(u) = e^{\jj \frac{2\pi n}{M}}
\left(
\left[ \Re(\tilde{u}) \right]_{0}^{\cos(\pi/M)} + \jj \left[ \Im(\tilde{u}) \right]_{-\sin(\pi/M)}^{\sin(\pi/M)}
\right),
\end{equation}
where
\[
n = \left\lfloor  \tfrac{\angle u + \pi/M }{2\pi/M} \right\rfloor, \quad
\tilde{u} = u e^{-\jj \frac{2\pi n}{M} }.
\]
In fact, one can even see the solution \eqref{eq:DCE_round} by pictures; see Fig.~\ref{fig:QCEproj} for one such picture.
\ifconfver
\begin{figure}
  \centering
  \includegraphics[width=0.6\linewidth]{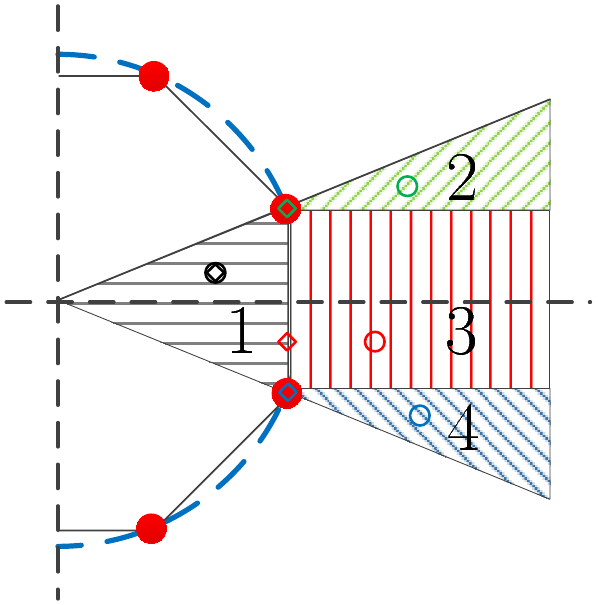}
%  \captionsetup{justification=centering,margin=2cm}
  \caption{
  Illustration of the projection onto $\setbarU$ for the DCE case.
  $M= 8$, the circle ``$\circ$'' and the diamond ``$\diamond$'' represent a given point $u$ and its projection $\Pi_{\setbarU}(u)$, resp.
  It can be seen that i) for $u$ lying in region 1, $\Pi_{\setbarU}(u)=u$;
  ii) for $u$ lying in region 2, $\Pi_{\bar{\setU}}(u)= \cos(\pi/M) + \jj \sin(\pi/M)$;
  iii) for $u$ lying in region 3, $\Pi_{\bar{\setU}}(u)= \cos(\pi/M) + \jj \Im(u)$;
  iv) for $u$ lying in region 4, $\Pi_{\bar{\setU}}(u)= \cos(\pi/M) - \jj \sin(\pi/M)$.}
  \label{fig:QCEproj}
\end{figure}
\else
\begin{figure}
  \centering
  \includegraphics[width=0.4\linewidth]{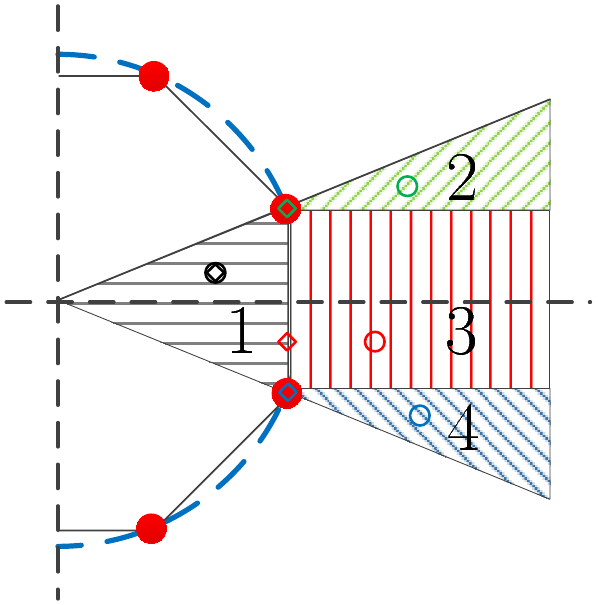}
%  \captionsetup{justification=centering,margin=2cm}
  \caption{
  Illustration of the projection onto $\setbarU$ for the DCE case.
  $M= 8$, the circle ``$\circ$'' and the diamond ``$\diamond$'' represent a given point $u$ and its projection $\Pi_{\setbarU}(u)$, resp.
  It can be seen that i) for $u$ lying in region 1, $\Pi_{\setbarU}(u)=u$;
  ii) for $u$ lying in region 2, $\Pi_{\bar{\setU}}(u)= \cos(\pi/M) + \jj \sin(\pi/M)$;
  iii) for $u$ lying in region 3, $\Pi_{\bar{\setU}}(u)= \cos(\pi/M) + \jj \Im(u)$;
  iv) for $u$ lying in region 4, $\Pi_{\bar{\setU}}(u)= \cos(\pi/M) - \jj \sin(\pi/M)$.}
  \label{fig:QCEproj}
\end{figure}
\fi
\ifconfver
The second key operation is with the computations of the gradient of $F_\lambda$, which is shown in \eqref{eq:grad_u} and \eqref{eq:grad_d} on the top of page 9; $\nabla_{\dIi} G_{\lambda}(\bU,\bd|\bar{\bU})$ takes the same form as
%$\nabla_{\dRi} G_{\lambda}(\bU,\bd|\bar{\bU})$
\eqref{eq:grad_d},
 with all ``$\Re$" and ``$R$" replaced with ``$\Im$" and ``$I$", resp.
\begin{figure*}[!bt]
  \normalsize
\begin{equation}\label{eq:grad_u}
\nabla_{\bu_t} G_{\lambda}(\bU,\bd|\bar{\bU}) =\sigma\frac{\sum_{i=1}^{K}\nabla_{\Re(\bu_t)}f_{i,t}(\bu_t,\bd) + \jj\sum_{i=1}^{K}\nabla_{\Im(\bu_t)}f_{i,t}(\bu_t,\bd)}{\sum_{t=1}^{T} \sum_{i=1}^{K}f_{i,t}(\bu_t,\bd)}-2\lambda \bar{\bu}_t,~~t=1,\ldots,T,
\end{equation}
\begin{equation}\label{eq:grad_d}
\nabla_{\dRi} G_{\lambda}(\bU,\bd|\bar{\bU})=\frac{-e^{-\frac{\bRi}{\sigma}}(1+\Re(s_{i,t}))+ e^{-\frac{\cRi}{\sigma}} (\Re(s_{i,t})-1)}{\sum_{t=1}^{T} \sum_{i=1}^{K}f_{i,t}(\bu_t,\bd)}, ~~i =1,\ldots,K,
\end{equation}
where
\begin{equation*}
\begin{aligned}
\nabla_{\Re(\bu_t)}f_{i,t}(\bu_t,\bd) &=\frac{1}{\sigma}\sqrt{\frac{P}{N}}\left( \left(e^{-\frac{\bRi}{\sigma}}-  e^{-\frac{\cRi}{\sigma}}\right)\Re(\bh_i) +\left( e^{-\frac{\cIi}{\sigma}}-e^{-\frac{\bIi}{\sigma}}\right)\Im(\bh_i)\right), \\
\nabla_{\Im(\bu_t)}f_{i,t}(\bu_t,\bd) & =\frac{1}{\sigma}\sqrt{\frac{P}{N}} \left( \left(e^{-\frac{\bRi}{\sigma}}-  e^{-\frac{\cRi}{\sigma}}\right)\Im(\bh_i) +\left( e^{-\frac{\bIi}{\sigma}}-e^{-\frac{\cIi}{\sigma}}\right)\Re(\bh_i)\right).
\end{aligned}
\end{equation*}

\hrulefill
\end{figure*}

\else
The second key operation is with the computations of the gradient of $F_\lambda$.
We have
\begin{equation}\label{eq:grad_u}
\nabla_{\bu_t} G_{\lambda}(\bU,\bd|\bar{\bU}) =\sigma\frac{\sum_{i=1}^{K}\nabla_{\Re(\bu_t)}f_{i,t}(\bu_t,\bd) + \jj\sum_{i=1}^{K}\nabla_{\Im(\bu_t)}f_{i,t}(\bu_t,\bd)}{\sum_{t=1}^{T} \sum_{i=1}^{K}f_{i,t}(\bu_t,\bd)}-2\lambda \bar{\bu}_t,~~t=1,\ldots,T,
\end{equation}
where
\begin{equation*}
\begin{aligned}
\nabla_{\Re(\bu_t)}f_{i,t}(\bu_t,\bd) &=\frac{1}{\sigma}\sqrt{\frac{P}{N}}\left( \left(e^{-\frac{\bRi}{\sigma}}-  e^{-\frac{\cRi}{\sigma}}\right)\Re(\bh_i) +\left( e^{-\frac{\cIi}{\sigma}}-e^{-\frac{\bIi}{\sigma}}\right)\Im(\bh_i)\right), \\
\nabla_{\Im(\bu_t)}f_{i,t}(\bu_t,\bd) & =\frac{1}{\sigma}\sqrt{\frac{P}{N}} \left( \left(e^{-\frac{\bRi}{\sigma}}-  e^{-\frac{\cRi}{\sigma}}\right)\Im(\bh_i) +\left( e^{-\frac{\bIi}{\sigma}}-e^{-\frac{\cIi}{\sigma}}\right)\Re(\bh_i)\right).
\end{aligned}
\end{equation*}
Also,
\begin{equation}\label{eq:grad_d}
\nabla_{\dRi} G_{\lambda}(\bU,\bd|\bar{\bU})=\frac{-e^{-\frac{\bRi}{\sigma}}(1+\Re(s_{i,t}))+ e^{-\frac{\cRi}{\sigma}} (\Re(s_{i,t})-1)}{\sum_{t=1}^{T} \sum_{i=1}^{K}f_{i,t}(\bu_t,\bd)}, ~~i =1,\ldots,K,
\end{equation}
and $\nabla_{\dIi} G_{\lambda}(\bU,\bd|\bar{\bU})$ takes the same form as
%$\nabla_{\dRi} G_{\lambda}(\bU,\bd|\bar{\bU})$
\eqref{eq:grad_d},
 with all ``$\Re$" and ``$R$" replaced with ``$\Im$" and ``$I$", resp.
\fi

%\ifconfver
%\begin{figure*}[!bt]
%  \normalsize
%\begin{equation*}
%\begin{aligned}
%\nabla_{\Re(\bu_t)}f_{i,t}(\bu_t,\bd) &=\frac{1}{\sigma}\sqrt{\frac{P}{N}}\left( \left(e^{-\frac{\bRi}{\sigma}}-  e^{-\frac{\cRi}{\sigma}}\right)\Re(\bh_i) +\left( e^{-\frac{\cIi}{\sigma}}-e^{-\frac{\bIi}{\sigma}}\right)\Im(\bh_i)\right), \\
%\nabla_{\Im(\bu_t)}f_{i,t}(\bu_t,\bd) & =\frac{1}{\sigma}\sqrt{\frac{P}{N}} \left( \left(e^{-\frac{\bRi}{\sigma}}-  e^{-\frac{\cRi}{\sigma}}\right)\Im(\bh_i) +\left( e^{-\frac{\bIi}{\sigma}}-e^{-\frac{\cIi}{\sigma}}\right)\Re(\bh_i)\right).
%\end{aligned}
%\end{equation*}
%\end{figure*}
%\else

\begin{algorithm}[htb!]
	%\caption{ Beamformer Design with Statistical CSIT}
	\caption{GEMM for CE, one-bit and DCE Precoding}
	\begin{algorithmic}[1]
		\STATE
		{\bf given} a starting point $(\bU^0,\bd^0)$, an extrapolation sequence $\{ \alpha_k \}_{k \geq 0}$,
		an initial penalty $\lambda > 0$, a penalty threshold $\lambda_{\sf upp} > 0$,
		an integer $J$, $c > 1$, $\delta > 0$.
		%Initialize  $\bZ^{0}=\bZ^{-1}=\bU^0$, $\bw^{0}=\bw^{-1}=\bd^{0}$,  $\lambda$, $\xi_0=1$, integer $J$, and $\delta>1$
        %\STATE Find $\beta_k$ by backtracking line search.
		%\STATE Find the $\beta_k$ via backtracking line search \cite{Beck2009};
		\STATE $k= 0$.
		\STATE $\bZ_U^{-1} = \bZ^0_U = \bU^0$, $\bz_d^{-1} = \bz^0_d = \bd^0$.
        \REPEAT
		\STATE Update
		\begin{align*}
		\bZ^{k}_U&=\bU^{k}+\alpha_{k}(\bU^{k}-\bU^{k-1}), \\
		\bz^{k}_d&=\bd^{k}+\alpha_{k}(\bd^{k}-\bd^{k-1}).
		\end{align*}
		\STATE Find $\beta_k$ by backtracking line search.
		\STATE Update
		\begin{align*}
		\bU^{k+1} &=\Pi_{\setbarU^{N\times T}} \left({\bZ^k_U}-\frac{1}{\beta_k}\nabla_{\bU} G_{\lambda}(\bZ^k_U,\bz^k_d | \bU^{k})\right), \\
		\bd^{k+1} &=\Pi_{\setD}\left(\bz^k_d - \frac{1}{\beta_k}\nabla_{\bd} G_{\lambda}(\bZ^k_U,\bz^k_d | \bU^{k})\right).
		\end{align*}
%		\begin{subequations}
%			\begin{align}
%			% \bz^{l}&=p_{L_l}(\bw^{l});\\
%			\xi_{k}&=\frac{1+\sqrt{1+4 \xi_{k-1}^2}}{2}, \label{eq:alg_gemm_DCE_a}\\
%			\alpha_{k}&= \min\left\{\frac{\xi_{k-1}-1}{\xi_{k}}, \bar{\alpha} \right\}  \label{eq:alg_gemm_DCE_b}\\
%			\bZ^{k}&=\bU^{k}+\alpha_{k}(\bU^{k}-\bU^{k-1}),  \label{eq:alg_gemm_DCE_c}\\
%			\bw^{k}&=\bd^{k}+\alpha_{k}(\bd^{k}-\bd^{k-1}),  \label{eq:alg_gemm_DCE_d}\\
%			\bU^{k+1}&=\Pi_{\setbarU^{N\times T}} \left({\bZ^k}-\frac{1}{\beta_k}\nabla_{\bU} G_{\lambda}(\bZ^k,\bw^k|\bU^{k})\right), \label{eq:alg_gemm_DCE_e}\\
%			\bd^{k+1}&=\Pi_{\setD}\left(\bw^k-\frac{1}{\beta_k}\nabla_{\bd} G_{\lambda}(\bZ^k,\bw^k|\bU^{k})\right);  \label{eq:alg_gemm_DCE_f}
%			\end{align}
%		\end{subequations}
		\STATE Update $\lambda= \lambda c$ every $J$ iterations, or if $\| \bU^{k+1} - \bU^k \|^2 + \| \bd^{k+1} - \bd^k \|^2 \leq \delta$.
		\STATE  $k=k+1$.
		%    \State Update $\bl^{(m+1)}= \lfloor \frac{\bs  -\bH \bx_m^{(nT)}}{\tau} \rceil$, where $\lfloor \cdot \rceil$ is rounding operator to the nearest integer.
		\UNTIL $\lambda > \lambda_{\sf upp}$.
	\end{algorithmic}\label{Al:GEMM_DCE}
\end{algorithm}

%------------------------------------------------
\section{Simulation Results}
\label{sec:simulation}

In this section, we illustrate the performance of our precoding design via Monte-Carlo simulations.

\subsection{One-Bit Precoding}

Firstly, we consider one-bit precoding.
The simulation settings are as follows.
We evaluate the average bit error rates (BERs) of our algorithms and some other algorithms over $10,000$ channel trials.
The channels $\bh_i$'s at each trial are randomly generated,
and we use the standard circular complex Gaussian distribution to generate the elements of $\bh_i$'s in an independent and identical fashion.
The transmit power is set to $P=1$.
We benchmark our algorithm against several other algorithms.
The first is the zero-forcing (ZF) precoder
\[
\bm\xi_t^{\sf ZF} = d \bH^H (\bH\bH^H)^{-1} \bs_t, \quad t=1,\ldots,T,
\]
where $d$ is chosen such that $\Exp_{\bs_t}[ \| \bm\xi_t^{\sf ZF} \|^2 ] = P$.
%This ZF precoder is a free-space precoder, or a precoder without one-bit constraints.
The reason for including the ZF precoder in our simulations is to help us evaluate how close a one-bit precoder can achieve compared to a free-space precoder.
The second is the quantized ZF (QZF) precoder, where we element-wise quantize $\bm\xi_t^{\sf ZF}$ to the nearest point in $\setU_{\sf 1-bit}$
and use that as the one-bit precoder.
The third is the SQUID algorithm proposed in \cite{Jacobsson2016}, which is an MMSE-based design.
Following the original work \cite{Jacobsson2017,Jacobsson2016}, SQUID is implemented by the Douglas-Rachford splitting method with the maximum number of iterations  set to $50$.
 The fourth is the iterative discrete estimation (IDE) method proposed in \cite{wang2018finite}, which adopts the same MMSE-based design as SQUID but uses a different optimization algorithm.
% The original IDE method considers precoding design in a per-symbol basis; herein we extend it to the block precoding case.
The fifth is the multi-user transmitting signal design (MUTSD) proposed in \cite{Sohrabi2018}, which is an SEP-based design but uses a different design formulation from ours.

The settings of our algorithm, GEMM, are as follows.
The smoothing parameter is $\sigma= 0.05$;
the penalty parameter is initialized as $\lambda = 0.01$;
it is increased by a factor of $c= 5$ when the number of GEMM iterations is more than $J=400$ or when the distance of successive iterates is less than $\delta= 10^{-4}$,
and the algorithm stops when $\lambda > 100$;
 and we initialize the algorithm by random initialization.
By our numerical experience, we found that GEMM is not too sensitive to initialization.
Our numerical experience also indicates that good results are generally yielded if we choose a small initial $\lambda$ and increase $\lambda$ gradually (which means a not too large $c$).
The intuition for such a parameter selection is that we may tackle the problem better if we gradually increase the hardness of the problem; note that our problem in \eqref{eq:P_app} is convex when $\lambda = 0$, and concave (and undesirable) for sufficiently large $\lambda$.
Furthermore, the smoothing parameter $\sigma$ should not be too small.
Naturally we desire to have $\sigma$ as small as possible, but
reducing $\sigma$ also increases the Lipschitz constant of the gradient of the objective function, which can lead to slower convergence (cf., Theorem~\ref{thm:GEMM_conv}).

In addition to GEMM,
we also try MM, or more precisely, the exact MM with the APG method as the solver for the MM iterations.

Figs.~\ref{fig:OB16} and \ref{fig:OB64} show the BERs for the $16$-QAM and $64$-QAM cases, resp., and for $(N,K,T) = (128,16,10)$.
It is seen that GEMM and MM perform better than SQUID and QZF.
%The reason for this may be that GEMM and MM are SEP-based designs, while SQUID and QZF are not.
Also, for the $16$-QAM case, the SNR gap between ZF and GEMM (or MM) is about $5$dB.
This number is encouraging as it suggests that one-bit precoding has the potential of offering comparable performance relative to free-space ZF precoding.
However, the situation is not as promising for the $64$-QAM case, where the SNR gap is widened to more than $10$dB.
But still, the performance of GEMM and MM is reasonably good and does not show error floor effects as in SQUID and QZF.
\ifconfver
\begin{figure}[htb!]
	\centering
	\includegraphics[width=\linewidth]{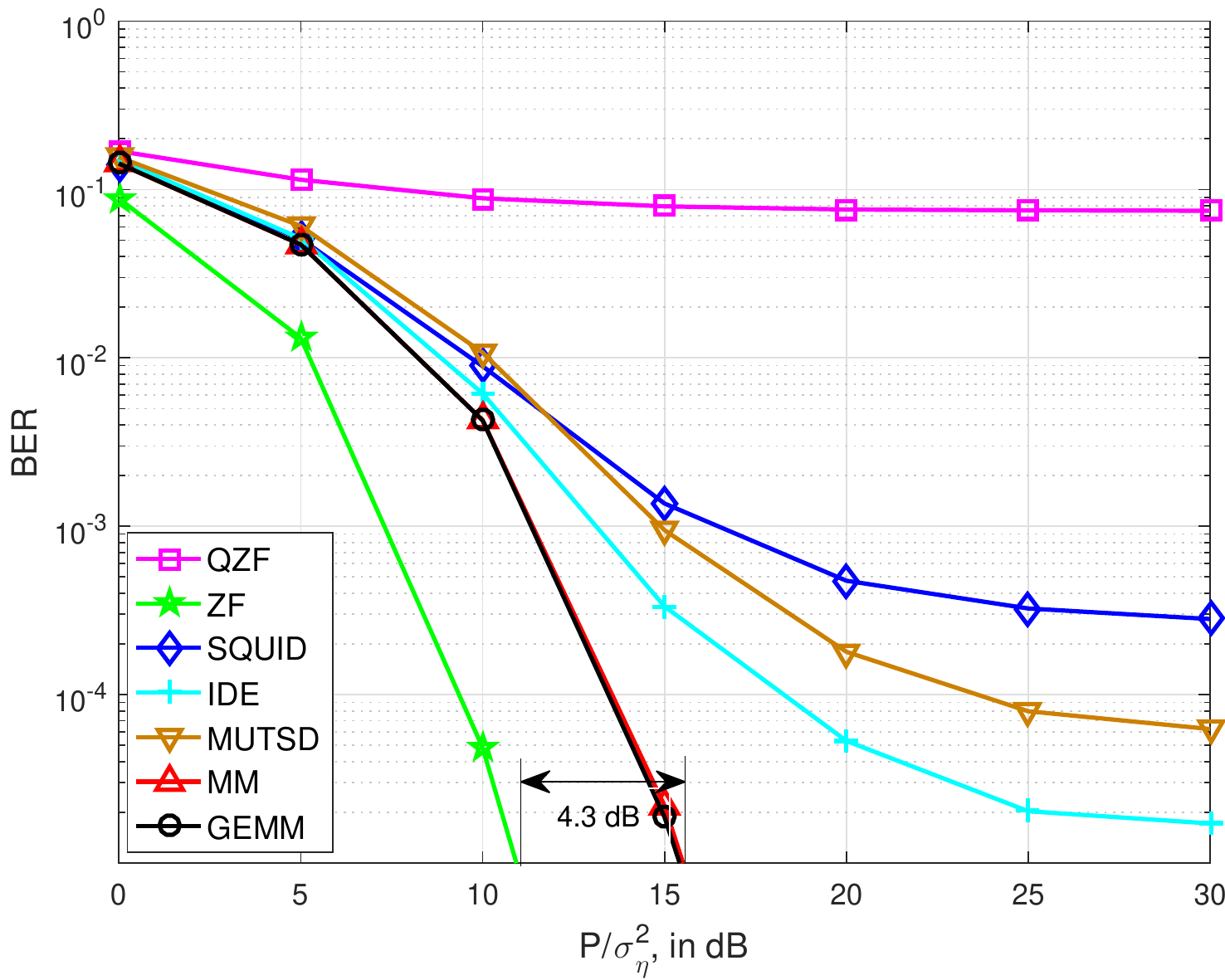}
	\caption{BER performance for one-bit precoding. $(N,K,T)=(128,16,10)$, $16$-QAM.}\label{fig:OB16}
\end{figure}
\else
\begin{figure}[htb!]
	\centering
	\includegraphics[width=0.6\linewidth]{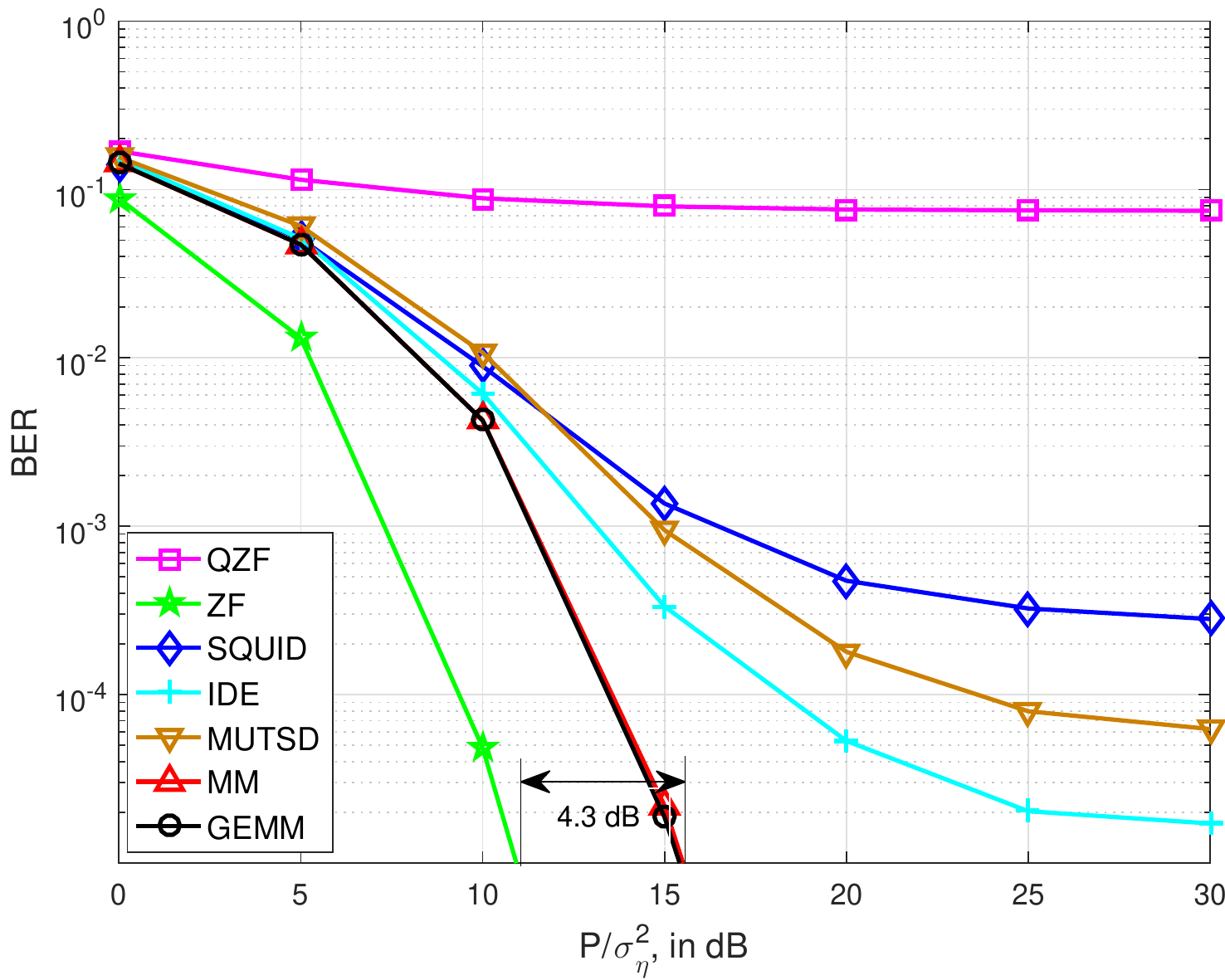}
	\caption{BER performance for one-bit precoding. $(N,K,T)=(128,16,10)$, $16$-QAM.}\label{fig:OB16}
\end{figure}
\fi

\ifconfver
\begin{figure}[htb!]
	\centering
	\includegraphics[width=\linewidth]{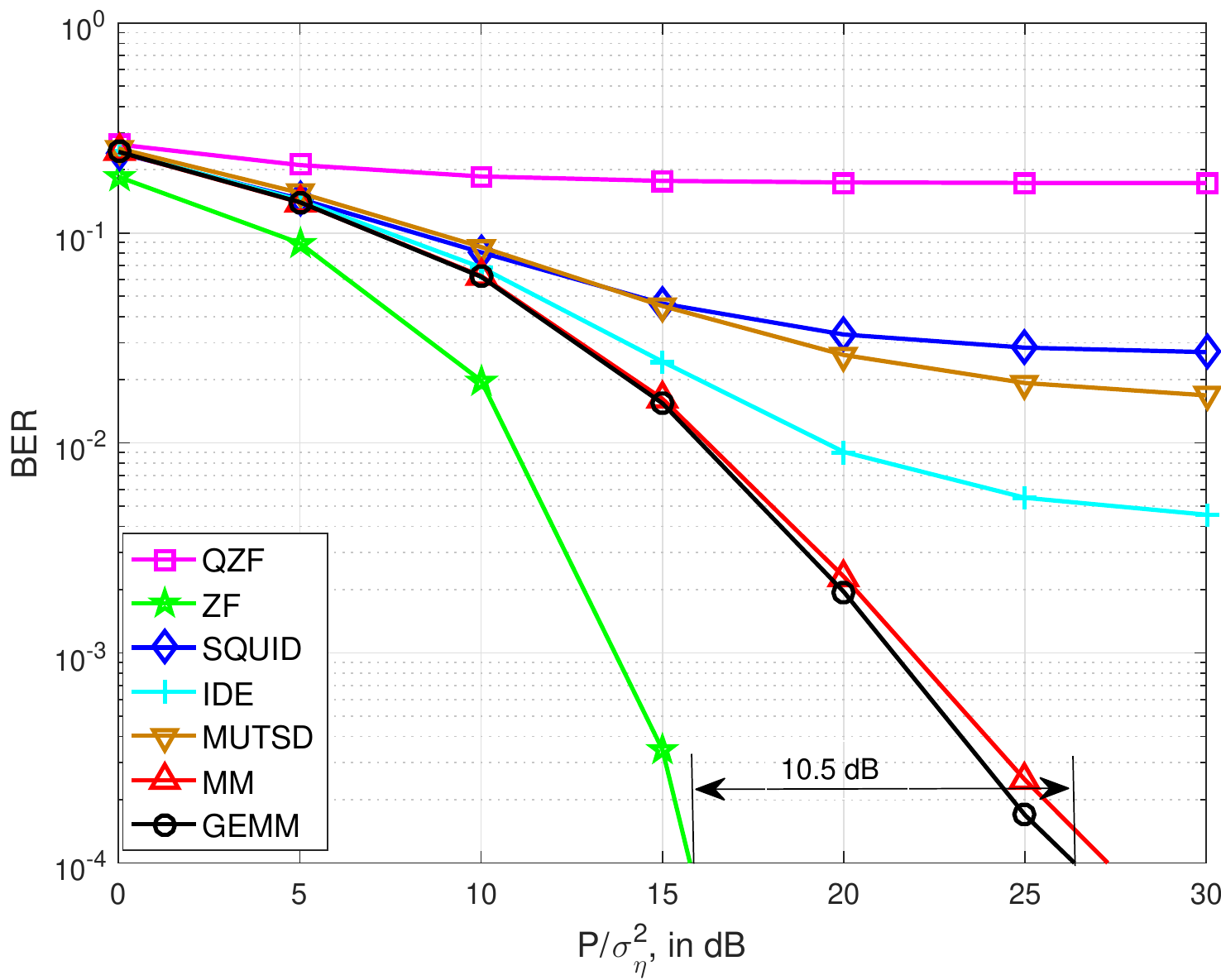}
	\caption{BER performance for one-bit precoding. $(N,K,T)=(128,16,10)$, $64$-QAM.}\label{fig:OB64}
\end{figure}
\else
\begin{figure}[htb!]
	\centering
	\includegraphics[width=0.6\linewidth]{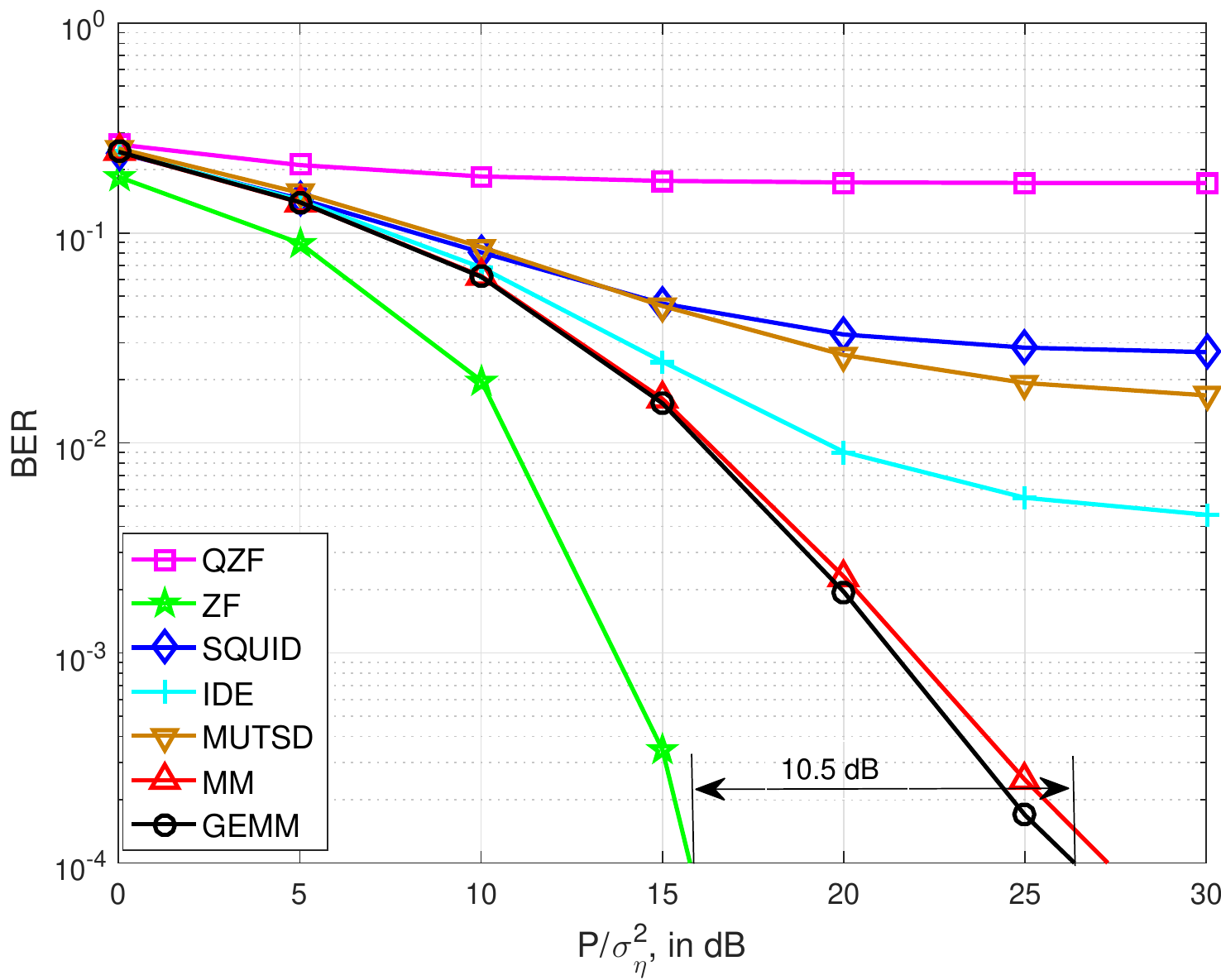}
	\caption{BER performance for one-bit precoding. $(N,K,T)=(128,16,10)$, $64$-QAM.}\label{fig:OB64}
\end{figure}
\fi

Table~\ref{tb:time} compares the runtimes of SQUID, MM and GEMM.
In this simulation we set $K= 16$, $T= 10$, and the QAM size to be $64$.
The simulation was conducted by MATLAB on a desktop computer with Intel i7-4770 processor and $16$GB memory.
We observe that GEMM is the fastest for larger problem dimensions, specifically, $N=256$ and $N=512$, and that the runtime differences between GEMM and
%SQUID, IDE and MUTSD
the other algorithms
are significant when $N$ increases.
GEMM is also seen to run about three times faster than MM.
\ifconfver
 \begin{table}[htb!]
	\centering
	\captionsetup{justification=centering}
	\caption{Average runtime (in Sec.) for each transmission block; $(K,T)=(16,10)$, $64$-QAM, one-bit precoding. \\[-0.4em]}\label{tb:time}
	\renewcommand{\arraystretch}{1.2}
	\resizebox{\linewidth}{!}{%
		\begin{tabular}{M{20mm}|M{10mm} M{10mm} M{10mm} M{10mm} }
			\hline
			$N$ & 128 & 192 & 256 & 512\\ \hline\hline
			SQUID  & 3.06 & 6.67 & 11.91 & 55.03 \\
            IDE  &{\bf 0.16} & {\bf 0.31} & 0.63 &3.83\\
            MUTSD  & 2.95 &3.23 & 3.54 & 5.46\\
			MM  & 1.35 & 1.46 & 1.51 & 2.17 \\
			GEMM  & 0.36 &  0.41 &  {\bf  0.49} & {\bf 0.85}\\ \hline
		\end{tabular}
	}
\end{table}
\else
 \begin{table}[htb!]
	\centering
	\captionsetup{justification=centering}
	\caption{Average runtime (in Sec.) for each transmission block; $(K,T)=(16,10)$, $64$-QAM\\[-0.4em]}\label{tb:time}
	\renewcommand{\arraystretch}{1.2}
	\resizebox{0.8\linewidth}{!}{%
		\begin{tabular}{M{30mm}|M{20mm} M{25mm} M{20mm} M{20mm} }
			\hline
			$N$ & 128 & 192 & 256 & 512\\ \hline\hline
			SQUID  & 3.06 & 6.67 & 11.91 & 55.03 \\
            IDE  &{\bf 0.16} & {\bf 0.31} & 0.63 &3.83\\
            MUTSD  & 2.95 &3.23 & 3.54 & 5.46\\
			MM  & 1.35 & 1.46 & 1.51 & 2.17 \\
			GEMM  & 0.36 &  0.41 &  {\bf  0.49} & {\bf 0.85}\\ \hline
		\end{tabular}
	}
\end{table}
\fi

In the previous BER simulation, we have chosen the transmission block length to be $T=10$.
In practice, the block length can be as large as a few hundreds.
Fig.~\ref{fig:OBlong} shows a BER result wherein not only the block length is increased to $T=200$, but we also scale up the number of transmit antennas and the number of users to $N=256$ and $K=24$, resp.
This results in a design problem whose number of decision variables exceeds $100,000$, which is computationally challenging.
We found that SQUID cannot be run (at least with our computer).
However, MM and GEMM can still be run; e.g., GEMM took about $4$ second for each trial.
We see that MM and GEMM provide reasonably good performance,
%In fact, the performance is consistent with that
as
in the previous simulation in Figs.~\ref{fig:OB16} and \ref{fig:OB64}.

\ifconfver
\begin{figure}[htb!]
	\centering
	\includegraphics[width=\linewidth]{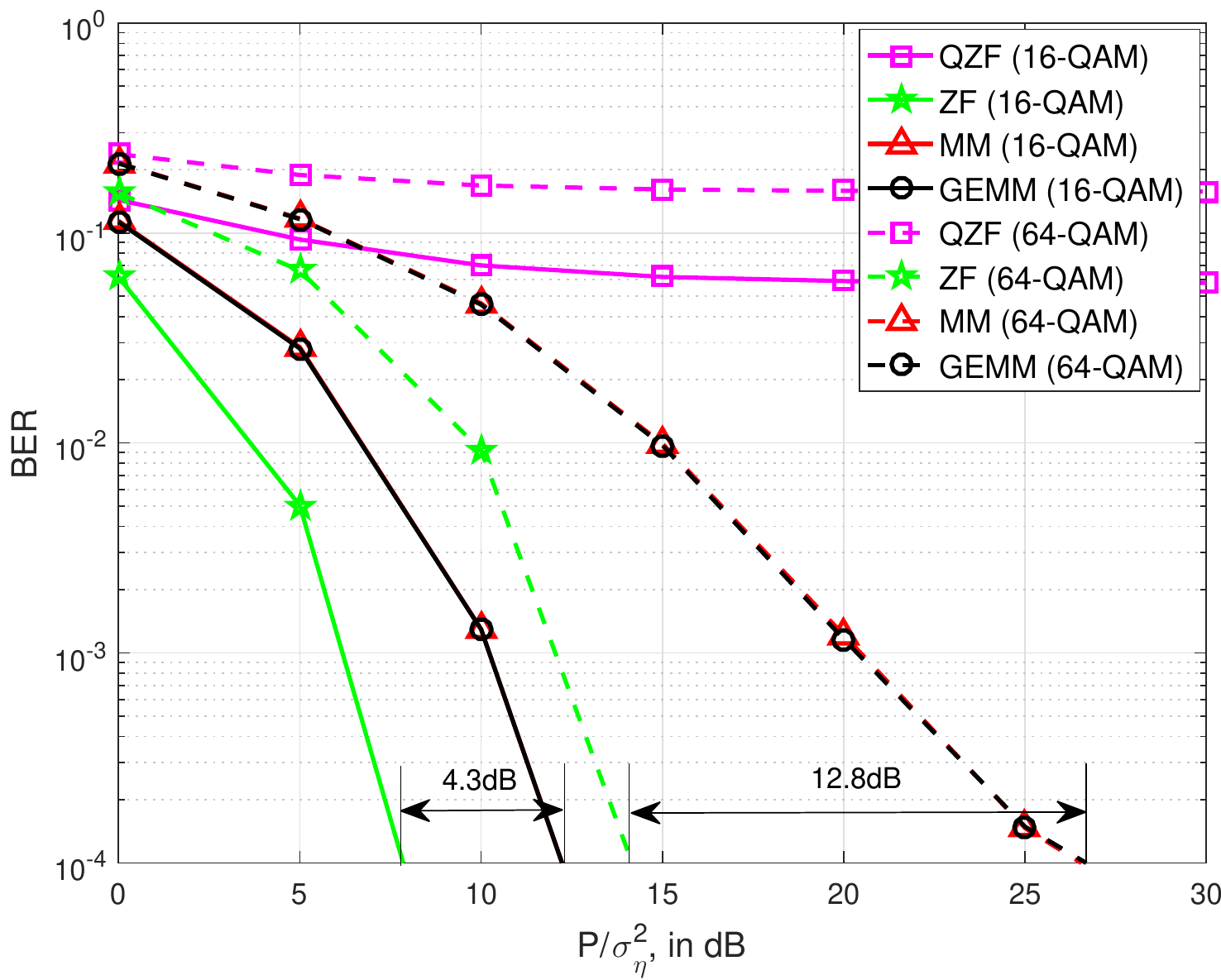}
	\caption{BER performance for one-bit precoding. $(N,K,T)=(256,24,200)$.}\label{fig:OBlong}
\end{figure}
\else
\begin{figure}[htb!]
	\centering
	\includegraphics[width=0.6\linewidth]{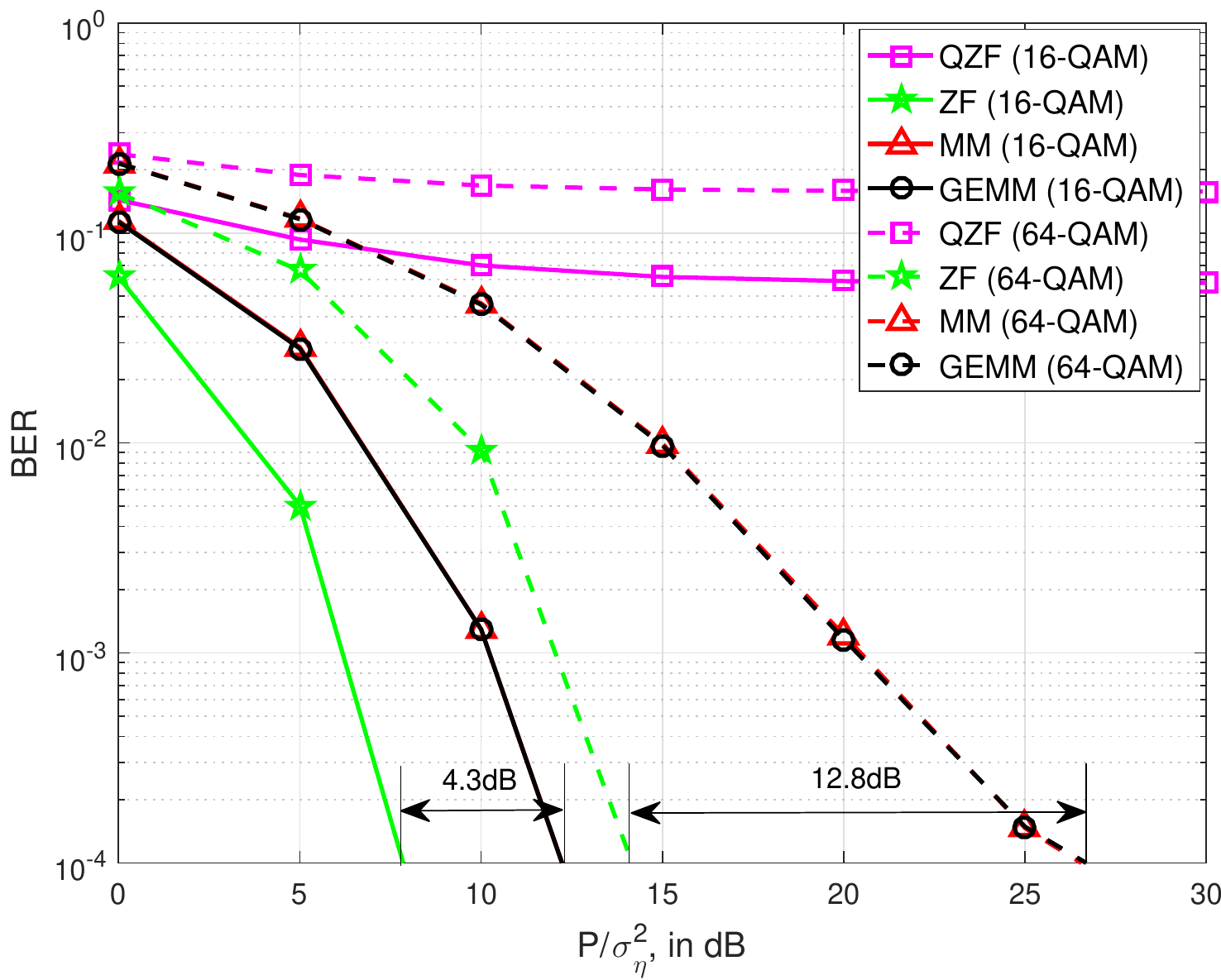}
	\caption{BER performance for one-bit precoding. $(N,K,T)=(256,24,200)$.}\label{fig:OBlong}
\end{figure}
\fi

%------------
\subsection{CE Precoding}

Secondly, we consider CE precoding.
The simulation settings are the same as those in the last subsection.
The benchmarked algorithms are ZF, QZF (with the quantization changed to that of the CE set), and an existing algorithm called MUImin~\cite{Mohammed2013}.
We no longer show the results for MM.
Like the results in the last subsection, we found that MM and GEMM provide almost the same BER performance, but GEMM runs faster than MM.

Fig.~\ref{fig:CE16} shows the BER results for $16$-QAM and $(N,K,T) = (128,16,10)$.
It is seen that GEMM performs better than MUImin and QZF,
and the SNR gap between GEMM and ZF is about $2$dB only---which, again, is promising.
Table~\ref{tb:CEtime} compares the average runtimes of GEMM and MUImin for $64$-QAM and $(K,T)=(16,10)$.
GEMM is seen to run faster than MUImin.
Fig. \ref{fig:CElong} shows the BERs of GEMM and ZF under different QAMs, specifically, $16$-QAM, $64$-QAM and $256$-QAM; we set $(N,K,T) = (128,16,50)$. It is interesting to see that the SNR gap between GEMM and ZF is within $5$dB even for  $256$-QAM;
again, such a result is encouraging.
\ifconfver
\begin{table}[htb!]
	\centering
	\captionsetup{justification=centering}
	\caption{Average runtime (in Sec.) for each transmission block; $(K,T)=(16,10)$, $64$-QAM, CE precoding. \\[-0.4em]}\label{tb:CEtime}
	\renewcommand{\arraystretch}{1.2}
	\resizebox{\linewidth}{!}{%
		\begin{tabular}{M{20mm}|M{10mm} M{10mm} M{10mm} M{10mm} }
			\hline
			$N$ & 64 & 128 & 192 & 256 \\ \hline\hline
			MUImin & 0.53 & 0.42 & 0.62 & 0.93 \\
			GEMM & {\bf 0.14} & {\bf 0.18} &{\bf 0.23} & {\bf 0.27} \\ \hline
			%NPG & 1.39 & 1.92 & 2.36 & 2.81 \\
			%ANPG & 0.72 & 1.01 & 1.45 & 1.67 \\
		\end{tabular}
	}
\end{table}
\else
\begin{table}[htb!]
	\centering
	\captionsetup{justification=centering}
	\caption{Average runtime (in Sec.) for each transmission block; $(K,T)=(16,10)$, $64$-QAM\\[-0.4em]}\label{tb:CEtime}
	\renewcommand{\arraystretch}{1.2}
	\resizebox{0.8\linewidth}{!}{%
		\begin{tabular}{M{30mm}|M{20mm} M{25mm} M{20mm} M{20mm} }
			\hline
			$N$ & 64 & 128 & 192 & 256 \\ \hline\hline
			MUImin & 0.53 & 0.42 & 0.62 & 0.93 \\
			GEMM & {\bf 0.14} & {\bf 0.18} &{\bf 0.23} & {\bf 0.27} \\ \hline
			%NPG & 1.39 & 1.92 & 2.36 & 2.81 \\
			%ANPG & 0.72 & 1.01 & 1.45 & 1.67 \\
		\end{tabular}
	}
\end{table}
\fi

%The result demonstrates that CE precoding is capable of offering comparable performance relative to free-space ZF precoding.

\ifconfver
\begin{figure}[htb!]
	\centering
	\includegraphics[width=\linewidth]{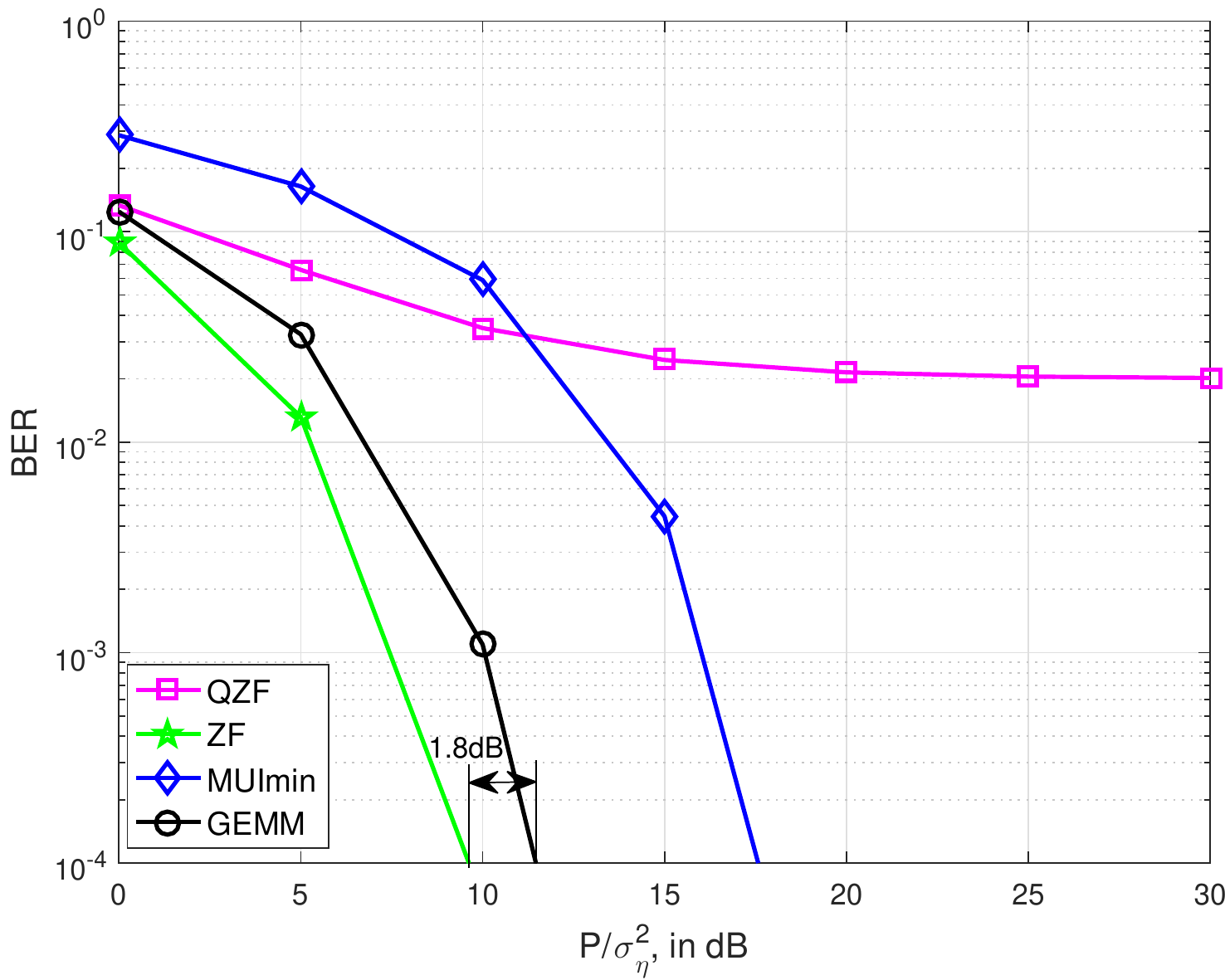}
	\caption{BER performance for CE precoding. $(N,K,T)=(128,16,10)$, 16-QAM.}\label{fig:CE16}
\end{figure}
\else
\begin{figure}[htb!]
	\centering
	\includegraphics[width=0.6\linewidth]{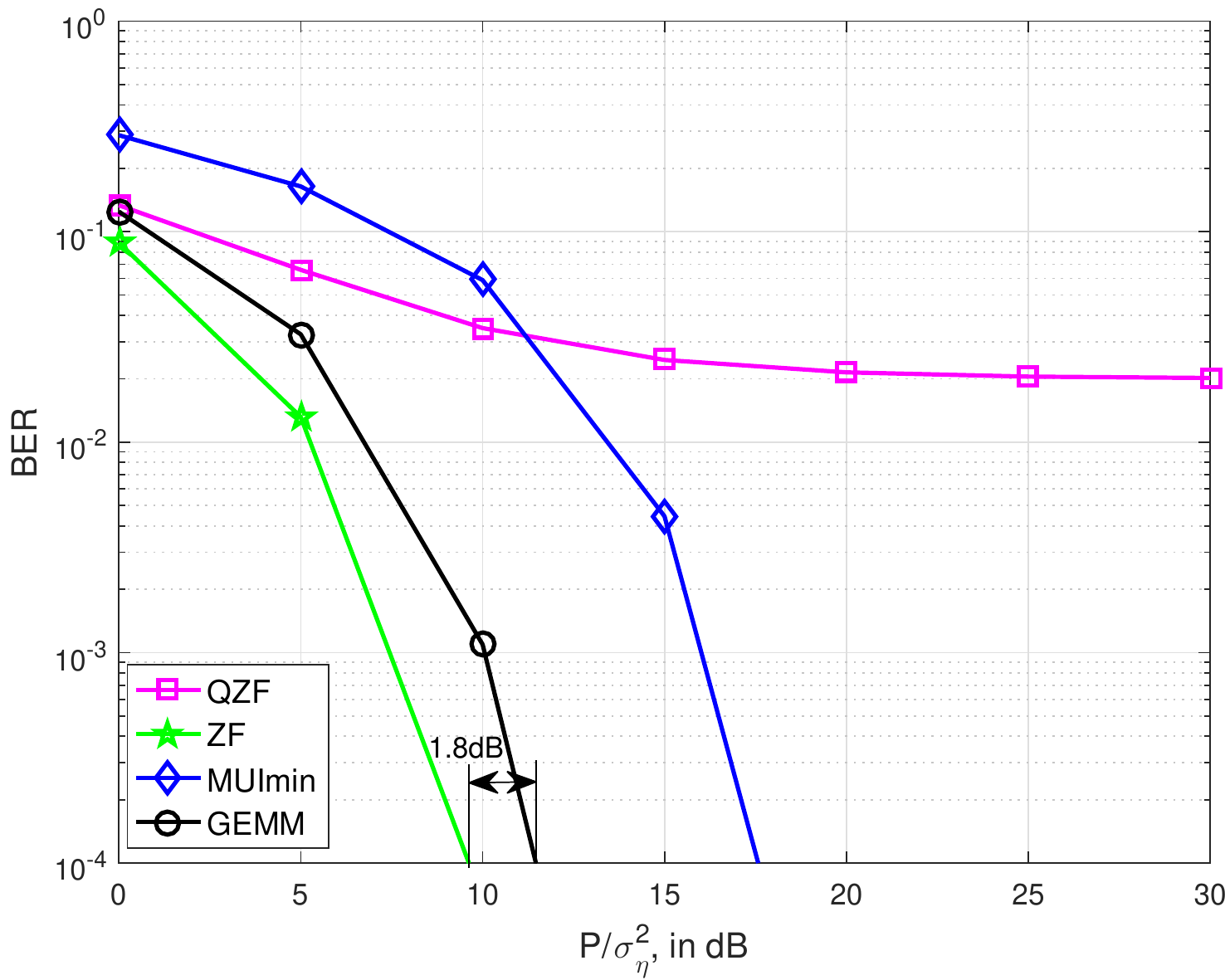}
	\caption{BER performance for CE precoding. $(N,K,T)=(128,16,10)$, 16-QAM.}\label{fig:CE16}
\end{figure}
\fi

\ifconfver
\begin{figure}[htb!]
	\centering
	\includegraphics[width=\linewidth]{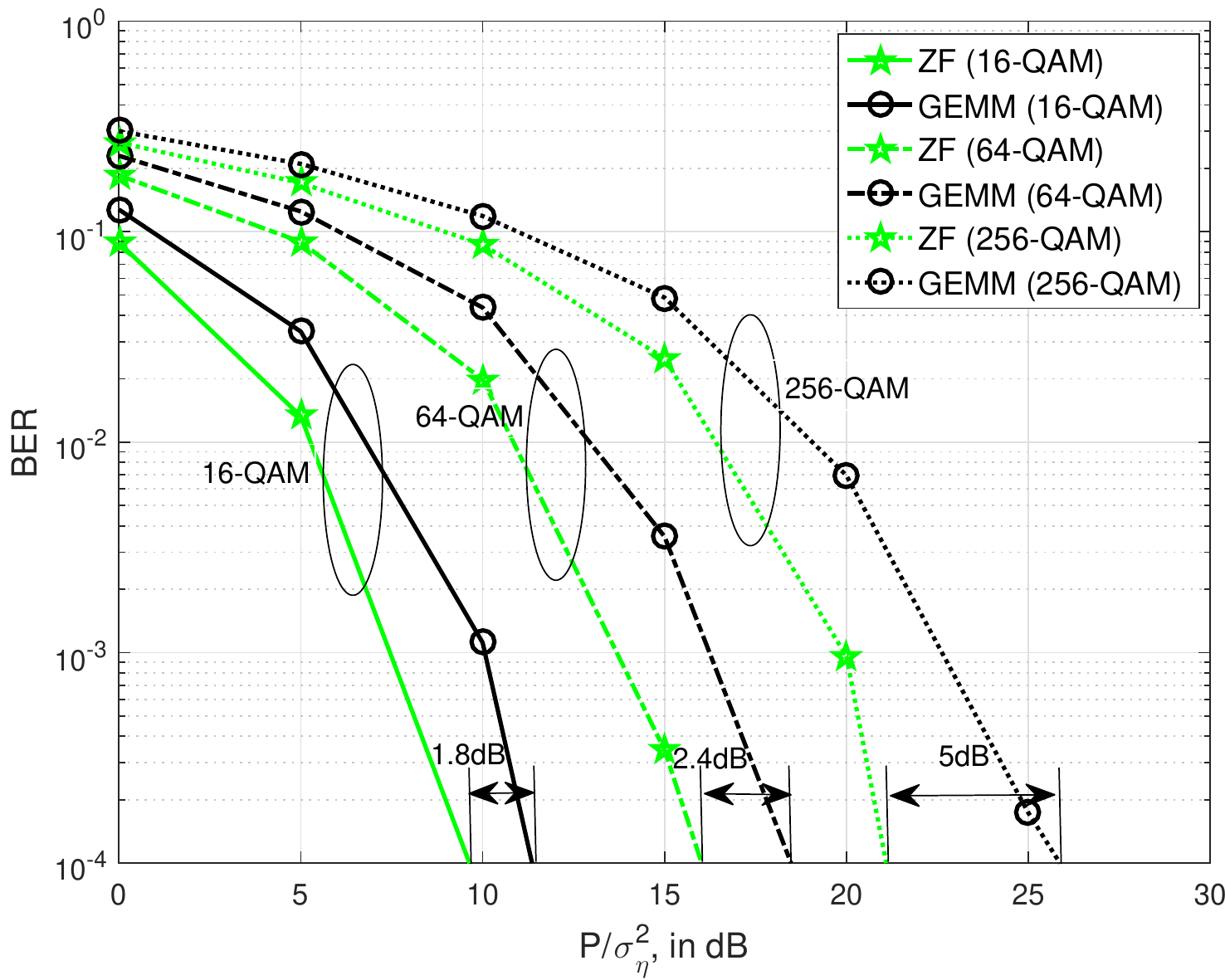}
	\caption{BER performance for CE precoding. $(N,K,T)=(128,16,50)$ with different QAM sizes.}\label{fig:CElong}
\end{figure}
\else
\begin{figure}[htb!]
	\centering
	\includegraphics[width=0.6\linewidth]{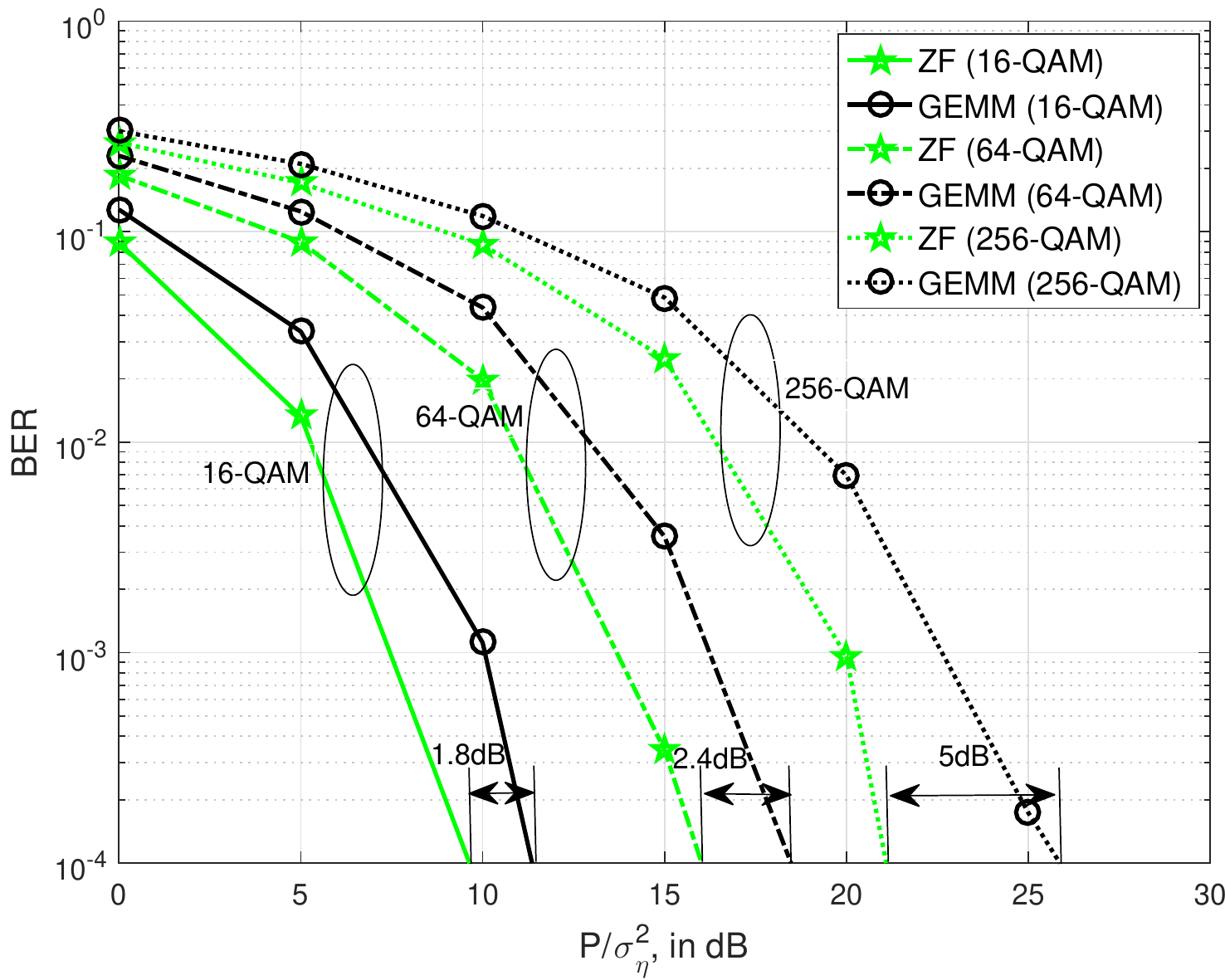}
	\caption{BER performance for CE precoding. $(N,K,T)=(128,16,50)$ with different QAM sizes.}\label{fig:CElong}
\end{figure}
\fi

%------------
\subsection{DCE Precoding}

Finally, we consider DCE precoding.
Fig.~\ref{fig:DCE} shows BER results for $64$-QAM, $(N,K,T) = (128,16,100)$ and under different numbers of phase combinations $M$.
%In this simulation we also try the following alternative design: We first use our (continuous) CE precoding design to obtain a CE design, and then we round the CE design to form a DCE design.
%%The reason for trying this alternative is to see under which conditions one may simply use a naively discrete CE precoder to perform DCE precoding.
%The aforementioned alternative is marked as ``CE Quant.'' in Fig.~\ref{fig:DCE}.
%We observe that ``CE Quant.'' indeed gives bad BER performance for $M= 4$, compared to DCE precoding (or ``GEMM'' in the figure).
%However, for $M=16$, the BER performance of ``CE Quant.'' and DCE precoding are almost the same; the two BER curves overlap.
For benchmarking purposes we also plotted the CE precoding result, which appears as ``CE ($M=\infty$)'' in the figure.
We see that DCE precoding for $M= 8$ is about $2$dB away from CE precoding,
and that DCE precoding for $M=16$
%(as well as ``CE Quant.'' for $M=16$)
approaches the BER performance attained by CE precoding.
This suggests that DCE precoding with moderate phase resolutions has the potential of achieving near-CE precoding performance.

\ifconfver
\begin{figure}[htb!]
	\centering
	\includegraphics[width=\linewidth]{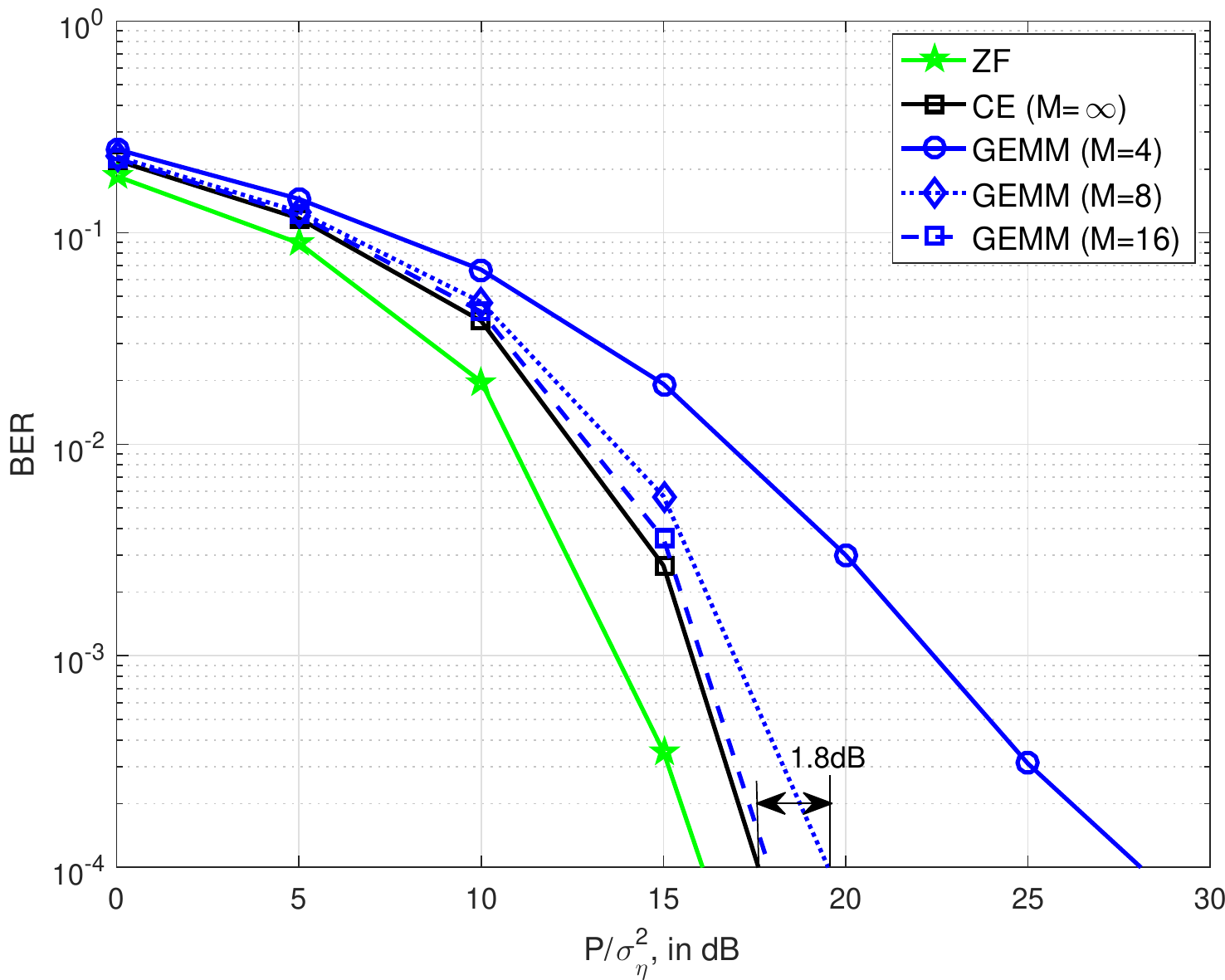}
	\caption{BER performance of DCE precoding.  $(N,K,T)=(128,16,100)$, $64$-QAM.}\label{fig:DCE}
\end{figure}
\else
\begin{figure}[htb!]
	\centering
	\includegraphics[width=0.6\linewidth]{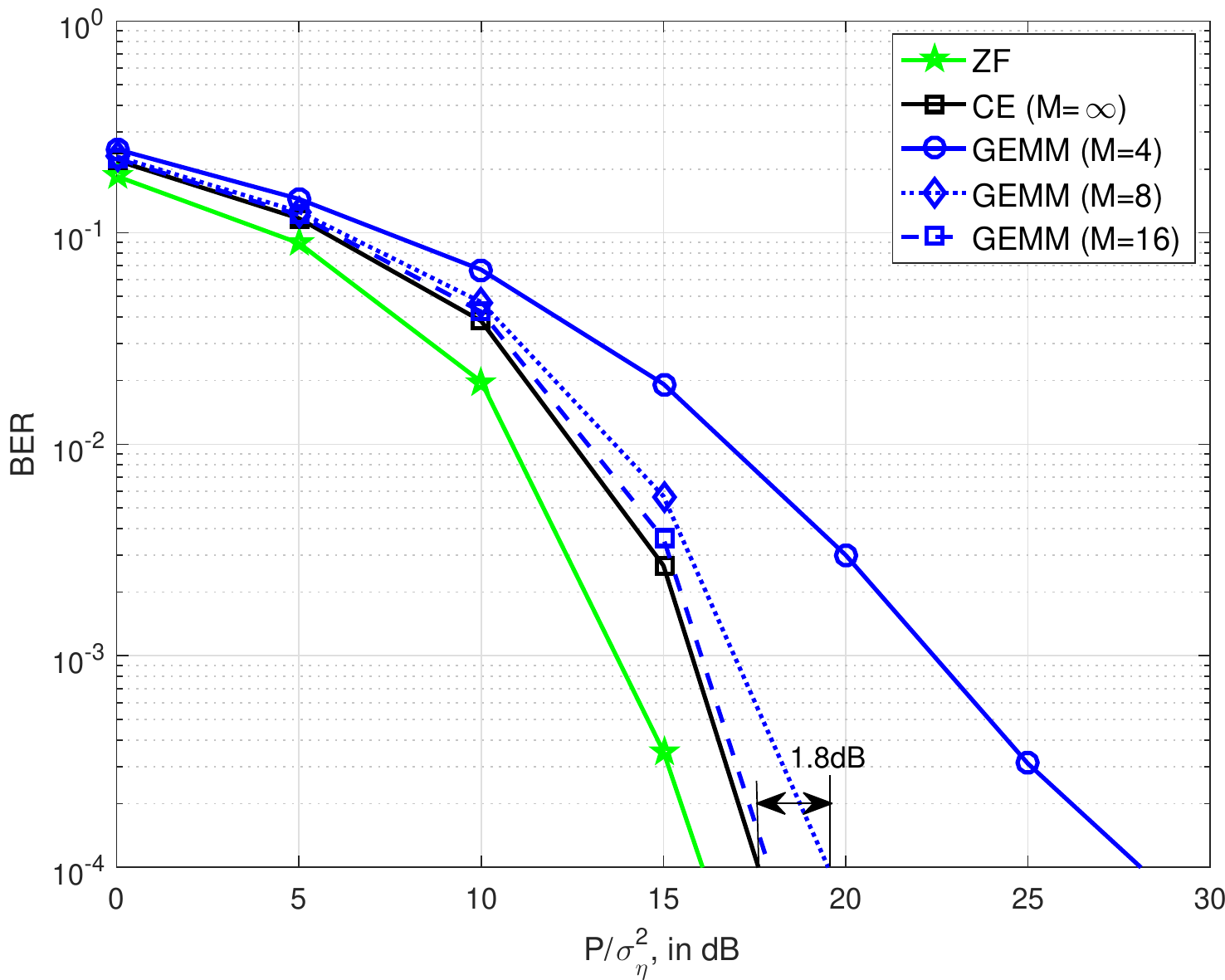}
	\caption{BER performance of DCE precoding.  $(N,K,T)=(128,16,100)$, $64$-QAM.}\label{fig:DCE}
\end{figure}
\fi

\section{Conclusion}\label{sec:cl}

In this paper we laid a framework for one-bit, CE and discrete CE precoding for the multiuser MISO downlink scenario.
The framework is SEP-based and focuses on optimization.
Simulation results indicated that the proposed framework provides satisfactory SEP performance; its runtime performance is also competitive.
We hope that this study would also provide a framework for attacking even more challenging precoding designs, such as the multiuser MIMO scenario, multi-bit precoding,
 and the scenario of imperfect channel information.

%-----------------------------------
%\vfill\pagebreak
%\newpage

\ifplainver
\section*{Appendix}
\renewcommand{\thesubsection}{\Alph{subsection}}
\else
\section{Appendix}
\fi

\subsection{Proof of Proposition~\ref{fact:d_bnd}}\label{appen:proof_d_bnd}

Let $(\bU^\star, \bd^\star)$ be an optimal solution to Problem~\eqref{eq:P1}.
Let $\hat{d}_i^R$ be an optimal solution to
\begin{align}
	& \min_{\dRi \geq 0 } ~
	\max_{  t= 1,\ldots, T} ~ \max\{ -\hat{b}_{i,t}^R, -\hat{c}_{i,t}^R \},
	\label{eq:proof_d_bnd_teq2a}
%	\\
%	& \min_{ \dIi \geq 0 } ~
%	\max_{ i=1,\ldots,K} ~ \max\{ -\hat{b}_{i,t}^I, -\hat{c}_{i,t}^I \},
%	\label{eq:proof_d_bnd_teq2b}
\end{align}
where $\hat{b}_{i,t}^R$ and $\hat{c}_{i,t}^R$ are given by \eqref{eq:bc} at $\bu_t = \bu_t^\star$, i.e.,
\begin{subequations}
	\label{eq:proof_d_bnd_teq11}
\begin{align}
\hat{b}_{i,t}^R & = (1 + \Re(s_{i,t}) )\dRi - \sqrt{\tfrac{P}{N}} \Re( \bh_i^T \bu_t^\star ), \\
\hat{c}_{i,t}^R & = (1- \Re(s_{i,t}) )\dRi + \sqrt{\tfrac{P}{N}} \Re( \bh_i^T \bu_t^\star ).
\end{align}
\end{subequations}
Similarly, let $\hat{d}_i^R$ be an optimal solution to
\begin{align}
	& \min_{ \dIi \geq 0 } ~
	\max_{ t= 1,\ldots, T} ~ \max\{ -\hat{b}_{i,t}^I, -\hat{c}_{i,t}^I \},
	\label{eq:proof_d_bnd_teq2b}
\end{align}
where $\hat{b}_{i,t}^I$ and $\hat{c}_{i,t}^I$ are obtained by changing ``$R$'' and ``$\Re$'' to ``$I$'' and ``$\Im$'', resp., in \eqref{eq:proof_d_bnd_teq11}.
Let $\hat{\bd}^R = [~ \hat{d}_1^R, \ldots, \hat{d}_K^R ~]^T$,
$\hat{\bd}^I = [~ \hat{d}_1^I, \ldots, \hat{d}_K^I ~]^T$,
$\hat{\bd} = [~ (\hat{\bd}^R)^T ~ (\hat{\bd}^I)^T  ]^T$.
First, we argue that $(\bU^\star, \hat{\bd})$ is also an optimal solution to Problem~\eqref{eq:P1}.
To see this,
consider fixing $\bU= \bU^\star$ in Problem~\eqref{eq:P1}; i.e.,
\begin{align}
& \min_{\substack{ \bd \geq \bzero }} ~
\max_{\substack{ i=1,\ldots,K, \\ t= 1,\ldots, T}} ~ \max\{ -\hat{b}_{i,t}^R, -\hat{c}_{i,t}^R, -\hat{b}_{i,t}^I, -\hat{c}_{i,t}^I \}.
\label{eq:proof_d_bnd_teq1}
\end{align}
Clearly,
if $\hat{\bd}$ is an optimal solution to Problem~\eqref{eq:proof_d_bnd_teq1},
then
$(\bU^\star,\hat{\bd})$ is an optimal solution to Problem~\eqref{eq:P1}.
By noting that $\hat{b}_{i,t}^R$ and $\hat{c}_{i,t}^R$ in \eqref{eq:proof_d_bnd_teq11} depend on $\dRi$ only,
and that the same applies when we change ``$R$'' to ``$I$'',
one can easily see that Problem~\eqref{eq:proof_d_bnd_teq1} can be decoupled as the problems in \eqref{eq:proof_d_bnd_teq2a} and \eqref{eq:proof_d_bnd_teq2b}.
Thus, $\hat{\bd}$ is an optimal solution to Problem~\eqref{eq:proof_d_bnd_teq1}.
%It follows that $(\bU^\star, \hat{\bd})$ is an optimal solution to Problem~\eqref{eq:P1}.

Second, we prove that there exist $\hat{d}_i^R$ and $\hat{d}_i^I$ such that they are bounded by $\sqrt{P/N} \| \bh_i \|_1$;
this, together with the optimality of $(\bU^\star, \hat{\bd})$, lead to Proposition~\ref{fact:d_bnd}.
The following lemma will be needed.

\begin{Lemma} \label{lem:pwl}
	Let
	\[
	f(x) = \max_{i=1,\ldots,m} a_i x + b_i,
	\]
	where $m \geq 2$ and $a_i \neq a_j$ for some $i \neq j$.
	Consider
	\begin{equation} \label{eq:proof_d_bnd_teq3}
	\min_x f(x),
	\end{equation}
	and suppose that $f$ is bounded below over $\Rbb$.
	Then, there exists an optimal solution $x^\star$ to Problem~\eqref{eq:proof_d_bnd_teq3} such that
	\begin{equation} \label{eq:proof_d_bnd_teq4}
	| x^\star | \leq \frac{\displaystyle \max_{j \neq k} | b_k - b_j | }{\displaystyle \min_{\substack{j \neq k, ~ a_j \neq a_k}} | a_j - a_k |  }.
	\end{equation}
\end{Lemma}

The proof of Lemma~\ref{lem:pwl} is shown in Appendix \ref{appen:proof_lm1}.
Let us consider $\hat{d}_i^R$.
%Since Problem~\eqref{eq:proof_d_bnd_teq2a} is a one-dimensional problem,
A solution $\hat{d}_i^R$ to Problem~\eqref{eq:proof_d_bnd_teq2a} is either $\hat{d}_i^R = 0$ or an unconstrained minimizer of the objective function of Problem~\eqref{eq:proof_d_bnd_teq2a}.
We can verify that the objective function of Problem~\eqref{eq:proof_d_bnd_teq2a} is bounded below on $\Rbb$.
Specifically, from \eqref{eq:proof_d_bnd_teq11} we have
\begin{align*}
\max\{ -\hat{b}_{i,t}^R, -\hat{c}_{i,t}^R \}  &
	= - \dRi + \left| \Re(s_{i,t}) \dRi  - \sqrt{\tfrac{P}{N}} \Re( \bh_i^T \bu_t^\star ) \right|  \\
	& \geq  - \dRi + \left| \Re(s_{i,t}) \right| |\dRi| - \left| \sqrt{\tfrac{P}{N}} \Re( \bh_i^T \bu_t^\star ) \right| \\
%	= & |\dRi| (\left| \Re(s_{i,t}) \right| - {\rm sgn}(\dRi)  )- \left| \sqrt{\tfrac{P}{N}} \Re( \bh_i^T \bu_t^\star ) \right| \\
	 & \geq - \left| \sqrt{\tfrac{P}{N}} \Re( \bh_i^T \bu_t^\star ) \right|,
\end{align*}
where the second inequality is due to $ \left| \Re(s_{i,t}) \right| \geq 1$ for all $s_{i,t} \in \setS$.
The above inequality implies that the objective function of Problem~\eqref{eq:proof_d_bnd_teq2a} is bounded below by $- \max_{t=1,\ldots,T} | \sqrt{P/N} \Re( \bh_i^T \bu_t^\star ) |$.
By applying Lemma~\ref{lem:pwl}, there exists an unconstrained minimizer $\hat{d}_i^R$ of the objective function of Problem~\eqref{eq:proof_d_bnd_teq2a} such that
\begin{align} \label{eq:proof_d_bnd_teq7}
|\hat{d}_i^R| & \leq \frac{  \sqrt{\tfrac{P}{N}} \max\{ C, D \} }{ \min\{ A, B \} },
\end{align}
where
\ifconfver
\begin{align*}
A & = \min_{\substack{ t \neq \tau, \\ \Re(s_{i,t}) \neq  \Re(s_{i,\tau}) }} | \Re(s_{i,t}) - \Re(s_{i,\tau})|, \\
 B &= \min_{\substack{ t, \tau, \\ \Re(s_{i,t}) \neq  -\Re(s_{i,\tau}) }} | \Re(s_{i,t}) + \Re(s_{i,\tau}) |,  \\
C & = \max_{ t \neq \tau} | \Re( \bh_i^T ( \bu_t^\star - \bu_\tau^\star )) |,
~~
D = \max_{ t, \tau} | \Re( \bh_i^T ( \bu_t^\star + \bu_\tau^\star )) |.
\end{align*}

\else
\begin{align*}
A & = \min_{\substack{ t \neq \tau, \\ \Re(s_{i,t}) \neq  \Re(s_{i,\tau}) }} | \Re(s_{i,t}) - \Re(s_{i,\tau})|, \quad B = \min_{\substack{ t, \tau, \\ \Re(s_{i,t}) \neq  -\Re(s_{i,\tau}) }} | \Re(s_{i,t}) + \Re(s_{i,\tau}) |,  \\
C & = \max_{ t \neq \tau} | \Re( \bh_i^T ( \bu_t^\star - \bu_\tau^\star )) |,
\quad
D = \max_{ t, \tau} | \Re( \bh_i^T ( \bu_t^\star + \bu_\tau^\star )) |.
\end{align*}

\fi
Notice that $A \geq 2, B \geq 2$ for any $s_{i,t}, s_{i,\tau} \in \setS$,
that $$C \leq \| \bh_i \|_1 \| \bu_t^\star - \bu_\tau^\star \|_\infty \leq 2 \| \bh_i \|_1,$$ and similarly, that $D \leq 2 \| \bh_i \|_1$.
Putting the above inequalities into \eqref{eq:proof_d_bnd_teq7}, and combining it with $\hat{d}_i^R \geq 0$, we are led to the final result $\hat{d}_i^R \leq \sqrt{P/N} \| \bh_i \|_1$.
Following the same proof as above, we also get $\hat{d}_i^I \leq \sqrt{P/N} \| \bh_i \|_1$.
The proof is complete.

\subsection{Proof of Lemma~\ref{lem:pwl}}\label{appen:proof_lm1}
To proceed, rewrite Problem~\eqref{eq:proof_d_bnd_teq3} as
\begin{equation}  \label{eq:proof_d_bnd_teq5}
\begin{aligned}
\min_{\bz} & ~ \bc^T \bz \\
{\rm s.t.} & ~ \bz \in \setP,
\end{aligned}
\end{equation}
where $\bz = [~ x, t ~]^T$, $\bc = [~ 0, 1 ~]^T$,
\begin{equation} \label{eq:proof_d_bnd_teq6}
\setP = \{ \bz \mid \ba_i^T \bz + b_i \leq 0, ~ i=1,\ldots,m \},
\end{equation}
and $\ba_i = [~ a_i, -1 ~]^T$, $i=1,\ldots,m$.
In particular, if $\bz^\star = [~ x^\star, t^\star ~]^T$ is an optimal solution to Problem~\eqref{eq:proof_d_bnd_teq5}, then $x^\star$ is an optimal solution to Problem~\eqref{eq:proof_d_bnd_teq3}, and $t^\star$ attains $t^\star = f(x^\star)$.
Firstly, we claim that $\setP$ has an extreme point.
A polyhedron in the form of \eqref{eq:proof_d_bnd_teq6} and with vector size $n$ is known to have an extreme point if and only if $\{ \ba_1,\ldots,\ba_m \}$ contains $n$ linearly independent vectors \cite[Proposition~2.1.5]{bertsekas2009convex}.
Since we have assumed that $a_i \neq a_j$ for some $i \neq j$, the corresponding $\ba_i, \ba_j$ are linearly independent.
It follows that $\{ \ba_1,\ldots,\ba_m \}$ contains $n=2$ linearly independent vectors,
and thus $\setP$ has an extreme point.

Secondly, we explore a relationship between the optimal solutions and extreme points.
It is known that if $\setP$ has an extreme point and $\bc^T \bz$ is bounded below over $\setP$,
then there exists an optimal solution $\bz^\star$ to Problem~\eqref{eq:proof_d_bnd_teq5} such that it is also an extreme point of $\setP$ \cite[Proposition~2.4.2]{bertsekas2009convex}.
We already showed that $\setP$ has an extreme point,
and it is easy to see that $\bc^T \bz$ is bounded below over $\setP$ if and only if $f$ is bounded below over $\Rbb$, which we assume.
As an extreme point of $\setP$, $\bz^\star$ satisfies the following condition \cite[Proposition~2.1.4]{bertsekas2009convex}:
the set $\{ \ba_i \mid \ba_i^T \bz^\star + b_i = 0 \}$ contains $n=2$ linearly independent vectors.
This implies that we can find two indices $j,k$ such that $\ba_j, \ba_k$ are linearly independent, and $\ba_j^T \bz^\star + b_j = \ba_k^T \bz^\star + b_k$.
The above equations are equivalent to
\[
a_j \neq a_k, \quad (a_j - a_k) x^\star  = b_k - b_j,
\]
and they imply that
\[
 \min_{j \neq k, a_j \neq a_k} |a_j - a_k| |x^\star| \leq \max_{j \neq k} | b_k - b_j |.
\]
The proof is complete.

\subsection{Proof of Theorem~\ref{thm:NEP}}
\label{appen:proof_thm_NEP}

Let us rewrite Problem~\eqref{eq:NEP_orig} as
\begin{equation} \label{eq:thm:NEP_t0}
F_{\sf orig}^\star  = \min_{\bu \in \setU^n} f(\bu) - \lambda \| \bu \|^2;
\end{equation}
note that $|u|= 1$ for any $u \in \setU$.
Also, denote
\begin{equation} \label{eq:thm:NEP_t1}
F_{\sf NSP}^\star = \min_{\bu \in \setbarU^n} f(\bu) - \lambda \| \bu \|^2.
\end{equation}
It is seen that $F_{\sf orig}^\star \geq F_{\sf NSP}^\star$.
Also, if all optimal solutions to Problem~\eqref{eq:thm:NEP_t1} lie in $\setU^n$,
then $F_{\sf orig}^\star =  F_{\sf NSP}^\star$ and the optimal solution sets of Problems~\eqref{eq:thm:NEP_t0}--\eqref{eq:thm:NEP_t1} are equivalent.
Now, we show that any optimal solution to Problems~\eqref{eq:thm:NEP_t1} must lie in $\setU^n$ if $\lambda \geq \bar{\lambda}$ ($\bar{\lambda}$ is defined in Theorem~\ref{thm:NEP}).
Let $\hat{\bu}$ be an optimal solution to Problem~\eqref{eq:thm:NEP_t1}, and suppose $\hat{\bu} \notin \setU^n$.
Then there exists an index $i$ such that $\hat{u}_i \notin \setU$.
Let $\hat{u}= \hat{u}_i$ and $h(u) = f(\hat{u}_1,\ldots,\hat{u}_{i-1},u,\hat{u}_{i+1},\ldots,\hat{u}_n)$.
We will show that there exists a  $\tilde{u} \in \setU$ such that
\begin{equation} \label{eq:thm:NEP_t2}
h(\hat{u}) - \lambda |\hat{u}|^2 > h(\tilde{u}) - \lambda |\tilde{u}|^2,
\end{equation}
which implies that $f(\bar{\bu}) - \lambda \| \bar{\bu} \|^2 > f({\bu}) - \lambda \| {\bu} \|^2$ for ${\bu} = (\hat{u}_1,\ldots,\hat{u}_{i-1},\tilde{u}_i,\hat{u}_{i+1},\ldots,\hat{u}_n)$, and which contradicts the optimality of $\hat{\bu}$ for  Problem~\eqref{eq:thm:NEP_t1}.

Firstly, consider the CE case for which $\setU = \{ u \mid |u|= 1 \}$ and $\setbarU = \{ u \mid |u| \leq 1 \}$.
Note $|\hat{u}| < 1$, and let $\tilde{u}= e^{\angle \hat{u}}$.
Also, let $L$ be a Lipschitz constant of $f$ on $\setbarU^n$, which, following the definition, is also a Lipschitz constant of $h$ on $\setbarU$.
We get
\begin{subequations} \label{eq:thm:NEP_t3}
	\begin{align}
		h(\hat{u}) - \lambda | \hat{u} |^2
			& \geq h(\tilde{u}) - L| \hat{u} - \tilde{u}| - \lambda |\hat{u}|
			\label{eq:thm:NEP_t3a} \\
			& = h(\tilde{u}) - \lambda |\tilde{u}|^2 + (\lambda - L)(1- | \hat{u}|),
			\label{eq:thm:NEP_t3b}
	\end{align}
\end{subequations}
where \eqref{eq:thm:NEP_t3a} is due to the Lipschitz continuity of $h$ and the fact that $a \geq a^2$ for $a \in [0,1]$;
\eqref{eq:thm:NEP_t3b} is due to $| \hat{u} - \tilde{u}| = 1 - |\hat{u}|$ and $|\tilde{u}|=1$.
Hence, if $\lambda > L$, then \eqref{eq:thm:NEP_t2} holds.

Secondly, consider the DCE case for which $\{ u = e^{\jj \left( \frac{2\pi}{M} m +  \frac{\pi}{M} \right) } \mid m=0,1,\ldots,M-1 \}$, with $M \geq 4$ and $M$ being even.
Note that the one-bit case is an instance of the DCE case, with $M=4$.
We divide the proof into two cases.
As the first case, suppose that $\hat{u}$ lies in the interior of $\setbarU$.
Let
\begin{equation} \label{eq:thm:NEP_t4}
r= \max \{ a \mid a e^{\angle \hat{u}} \in \setbarU \},
\end{equation}
and note $|\hat{u}| < r$.
Let us characterize $\hat{u}$ as $\hat{u} = \hat{\alpha} r e^{\angle \hat{u}}$, where $0 \leq \hat{\alpha} < 1$,
and let $\tilde{u}= r e^{\angle \hat{u}}$.
Following the same proof as in \eqref{eq:thm:NEP_t3}, we can readily show that
\[
h(\hat{u}) - \lambda | \hat{u} |^2
\geq h(\tilde{u}) - \lambda |\tilde{u}|^2 + (\lambda  r - L) r (1-  \hat{\alpha}).
\]
It follows that \eqref{eq:thm:NEP_t2} holds if $\lambda > L/r$.
As the second case, suppose that $\hat{u}$ lies in the boundary of $\setbarU$.
It can be seen, e.g., from Fig.~\ref{fig:convU}(c), that $\hat{u} \in \conv\{ v_1, v_2 \}$, where
\[
v_1 = e^{\jj \left( \frac{2\pi}{M} k +  \frac{\pi}{M} \right)}, \quad
v_2 = e^{\jj \left( \frac{2\pi}{M} (k+1) +  \frac{\pi}{M} \right)},
\]
for some integer $k$.
It can be shown that
\[
u \in \conv\{ v_1, v_2 \}
\quad \Longleftrightarrow \quad
u = \alpha d + c, ~ \alpha \in [-1,1],
\]
where
\begin{subequations}
	\begin{align}
	c & = \tfrac{1}{2} (v_1 + v_2 ) = e^{\jj  \frac{2\pi}{M} (k+1)} \cos(\pi/M),
	\label{eq:thm:NEP_t5a} \\
	d & = \tfrac{1}{2} (v_1 - v_2 ) = \jj e^{\jj  \frac{2\pi}{M} (k+1)} \sin(\pi/M).
	\label{eq:thm:NEP_t5b}
	\end{align}
\end{subequations}
Let us characterize $\hat{u}$ as $\hat{u}= \hat{\alpha} d + c$, where $|\hat{\alpha}| < 1$.
Also, let $\tilde{u}= d + c$ if $\hat{\alpha} \geq 0$ and $\tilde{u}= -d + c$ if
$\hat{\alpha} < 0$.
Following the same proof as in \eqref{eq:thm:NEP_t3}, we get
\begin{subequations}
	\begin{align}
	h(\hat{u})\! - \! \lambda | \hat{u} |^2 & \! \geq	
		h(\tilde{u}) - L | \hat{u} - \tilde{u}| - \lambda( |\hat{\alpha}|^2 |d|^2 + |c|^2)
	\label{eq:thm:NEP_t6a} \\
	 &  \geq	
	 h(\tilde{u}) - L | \hat{u} - \tilde{u}| - \lambda( |\hat{\alpha}| |d|^2 + |c|^2)
	\label{eq:thm:NEP_t6b} \\
	& = h(\tilde{u}) - \lambda | \tilde{u}|^2 \!+ \!(\lambda |d| - L ) |d| ( 1- |\hat{\alpha}|),
	\label{eq:thm:NEP_t6c}
	\end{align}
\end{subequations}
where we have used $| \alpha d + c |^2 = |\alpha|^2 |d|^2 + |c|^2$ in \eqref{eq:thm:NEP_t6a} and \eqref{eq:thm:NEP_t6c}.
Thus, \eqref{eq:thm:NEP_t2} holds if  $\lambda >L/|d|$.
Combining the above two cases, we further conclude that  \eqref{eq:thm:NEP_t2} holds if
\begin{equation} \label{eq:thm:NEP_t7}
\lambda >\max\{ L/|d|, L/r \}.
\end{equation}
It can be shown from \eqref{eq:thm:NEP_t4} that $r \geq \cos(\pi/M)$, and it is seen from \eqref{eq:thm:NEP_t5b} that $|d| = \sin(\pi/M)$.
Since $\cos(\pi/M) \geq \sin(\pi/M)$ for $M \geq 4$, \eqref{eq:thm:NEP_t7} is implied by $\lambda > L/\sin(\pi/M)$. The proof is complete.

\subsection{Proof of Theorem~\ref{theorem:NEP_smooth}} \label{app:proof_NEP_eqv_smooth}

Assume $\lambda > L/2$ throughout this proof.
Firstly, we show that $F_\lambda$ is strongly concave on $\setbarU^n$, i.e., there exists a constant $\alpha < 0$ such that
\begin{equation} \label{eq:thm2_proof_eq1}
\langle \nabla F_\lambda(\bu_1) - \nabla F_\lambda(\bu_2), \bu_1 - \bu_2 \rangle
%\langle \nabla g(\bu_1) - \nabla g(\bu_2), \bu_1 - \bu_2 \rangle
\leq \alpha \| \bu_1 - \bu_2 \|^2,
\end{equation}
for all $\bu_1, \bu_2 \in \setbarU^n$ with $\bu_1 \neq \bu_2$ \cite{bertsekas2003convex}.
The proof is as follows.
Since $\nabla F_\lambda(\bu) = \nabla f(\bu) - 2 \lambda \bu$, the left-hand side (LHS) of \eqref{eq:thm2_proof_eq1} equals
\begin{align*}
\text{LHS of \eqref{eq:thm2_proof_eq1}}
& = \langle \nabla f(\bu_1) - \nabla f(\bu_2), \bu_1 - \bu_2 \rangle - 2\lambda \| \bu_1 - \bu_2 \|^2    \\
& \leq L \| \bu_1 - \bu_2 \|^2 - 2 \lambda \| \bu_1 - \bu_2 \|^2,
\end{align*}
where the above inequality is due to the Cauchy-Schwartz inequality and the Lipschitz continuity of $\nabla f$ on $\setbarU^n$.
It follows that \eqref{eq:thm2_proof_eq1} holds with $\alpha = L - 2 \lambda$.

Secondly, we show that
any locally optimal solution to Problem~\eqref{eq:NEP} must be an extreme point of $\setbarU^n$, or equivalently, a point in $\setU^n$.
Let $\hat{\bu}$ be a locally optimal solution to Problem~\eqref{eq:NEP}.
By the definition of local optimality,
there exists a constant $\varepsilon > 0$ such that
\begin{equation} \label{eq:thm2_proof_eq2}
F_\lambda(\hat{\bu}) \leq F_\lambda(\bu), \quad \forall ~ \bu \in \setbarU^n \cap \mathcal{B}(\hat{\bu},\varepsilon),
\end{equation}
where $ \mathcal{B}(\hat{\bu},\varepsilon)= \{ \bu \in \Cbb^n \mid \| \bu - \hat{\bu} \| \leq \varepsilon \}$.
Suppose that $\hat{\bu}$ is not an extreme point of $\setbarU^n$.
This means that we can find $\bu_1,\bu_2 \in \setbarU^n$, with $\bu_1 \neq \hat{\bu}$, $\bu_2 \neq \hat{\bu}$, such that
\begin{equation} \label{eq:thm2_proof_eq3}
\hat{\bu} = \theta \bu_1 + (1-\theta) \bu_2,
\end{equation}
for some $\theta \in (0,1)$.
Let $\bv = \bu_1 - \bu_2$, and let
\begin{equation} \label{eq:thm2_proof_eq4}
\bar{\bu}_1 = \hat{\bu} - \alpha \bv, \quad
\bar{\bu}_2 = \hat{\bu} + \alpha \bv,
\end{equation}
for some $\alpha > 0$.
We argue that for a sufficiently small $\alpha$, it holds that $\bar{\bu}_1,\bar{\bu}_2 \in \setbarU^n \cap \mathcal{B}(\hat{\bu},\varepsilon)$.
It is immediate that $\bar{\bu}_1,\bar{\bu}_2 \in \mathcal{B}(\hat{\bu},\varepsilon)$ if $\alpha \leq \varepsilon/ \| \bv \|$.
To see why $\bar{\bu}_1,\bar{\bu}_2 \in \setbarU^n$, let $\mathcal{L} = \conv\{ \bu_1, \bu_2 \}$.
Since $ \setbarU^n$ is convex, we have $\mathcal{L} \subseteq  \setbarU^n$.
Also, by putting \eqref{eq:thm2_proof_eq3} into \eqref{eq:thm2_proof_eq4}, and noting $0 < \theta < 1$, one can verify that $\bar{\bu}_1,\bar{\bu}_2 \in \mathcal{L}$ whenever $\alpha \leq \min\{ \theta, 1 - \theta \}$.
Thus, we have  $\bar{\bu}_1,\bar{\bu}_2 \in \setbarU^n$ for $\alpha \leq \min\{ \theta, 1 - \theta \}$.
Now, by $\hat{\bu} = 0.5 \bar{\bu}_1 + 0.5 \bar{\bu}_2$ and the strong concavity of $F_\lambda$ on $\setbarU^n$, we get
\begin{align}
F_\lambda(\hat{\bu}) & > \tfrac{1}{2} F_\lambda(\bar{\bu}_1) + \tfrac{1}{2} F_\lambda(\bar{\bu}_2)
\nonumber \\
& \geq \min\{ F_\lambda(\bar{\bu}_1) , F_\lambda(\bar{\bu}_2) \}.
\label{eq:thm2_proof_eq5}
\end{align}
We see that \eqref{eq:thm2_proof_eq5} contradicts \eqref{eq:thm2_proof_eq2}.
Thus,
a locally optimal solution to Problem~\eqref{eq:NEP} must be an extreme point of $\setbarU^n$. The proof is complete.

%{\blue
%Furthermore, for CE case, we show any locally optimal solution $\hat{\bu}$ to \eqref{eq:NEP} is also a locally optimal solution \eqref{eq:NEP_orig}. From \eqref{eq:thm2_proof_eq2}, we have
%\begin{equation}\label{eq:local_min}
%  \begin{split}
%    f(\hat{\bu})=&F_{\lambda}(\hat{\bu})+\lambda \|\hat{\bu}\|^2\\
%    =& f(\hat{\bu})-\lambda \|\hat{\bu}\|^2+\lambda \|\hat{\bu}\|^2\\
%    \leq & f(\bu)-\lambda \|\bu\|^2+\lambda \|\bu^o\|^2, ~~ \forall \bu \in \mathcal{B}(\hat{\bu},\varepsilon) \bigcap \setU^n\\
%    = & f(\bu),~~ \forall \bu \in \mathcal{B}(\hat{\bu},\varepsilon) \bigcap \setU^n,
%  \end{split}
%\end{equation}
%where the last equation is due to $||\bu||^2=n$ for all $\bu \in \setU^n$. The result in \eqref{eq:local_min} implies that $\hat{\bu}$ is a local minimum of \eqref{eq:NEP_orig}.
%}

%Finally, the globally optimal solution equivalence can be easily seen
%Finally, we establish the global equivalence of Problems~\eqref{eq:NEP_orig} and \eqref{eq:NEP}. From the second step, we know that any globally optimal solution to Problem~\eqref{eq:NEP} is a feasible solution to Problem~\eqref{eq:NEP_orig}. On the other hand, since Problem~\eqref{eq:NEP} is a relaxation of Problem~\eqref{eq:NEP_orig}, any optimal solution to \eqref{eq:NEP_orig} is also
%a feasible solution to \eqref{eq:NEP}. Therefore, both Problems~\eqref{eq:NEP_orig} and \eqref{eq:NEP} have the same globally optimal solution set.
\subsection{Proof of Corollary~\ref{Cor:NEP}}
\label{appen:Cor:NEP}

Suppose $\lambda \geq L / \cos(\pi/M)$. Following the proof in
%the previous section,
Appendix~\ref{appen:proof_thm_NEP},
any optimal solution $\hat{\bu}$ to Problem~\eqref{eq:thm:NEP_t1} must satisfy
$\hat{u}_i \in \conv\{ v_{i,1}, v_{i,2} \}$ for all $i$,
where $v_{i,1} = e^{\jj \left( \frac{2\pi}{M} k_i +  \frac{\pi}{M} \right)},$
$v_{i,2} = e^{\jj \left( \frac{2\pi}{M} (k_i+1) +  \frac{\pi}{M} \right)}$
for some integer $k_i$.
For $i \in \{ 1,\ldots,n\}$, let
$\tilde{u}_i = v_{i,1}$ if $| \hat{u}_i - v_{i,1} | \leq | \hat{u}_i - v_{i,2} |$ and $\tilde{u}_i = v_{i,2}$ otherwise.
It can be verified that $\tilde{\bu} = \Pi_{\setU^n}(\hat{\bu})$, and that $| \hat{u}_i - \tilde{u}_i | \leq |d| = \sin(\pi/M)$.
It follows  from $F_{\sf orig}^\star \geq F_{\sf NSP}^\star$ that
\begin{align*}
F_{\sf orig}^\star & \geq F_{\sf NSP}^\star = f(\hat{\bu}) - \lambda \| \hat{\bu} \|^2 \\
& \geq f(\tilde{\bu}) - L \| \hat{\bu} - \tilde{\bu} \| - \lambda \| \hat{\bu} \|^2 \\
& \geq f(\tilde{\bu}) - L \sqrt{n} \sin(\pi/M) - \lambda n,
\end{align*}
which, in turn, implies $f^\star \geq f(\tilde{\bu}) - L \sqrt{n} \sin(\pi/M)$.
Also, the inequality $f^\star \leq f(\tilde{\bu})$ follows trivially from the fact that $\tilde{\bu} \in \setU^n$. The proof is complete.

\subsection{Proof of Fact \ref{fact:NEP_counterexample}}
\label{appen:NEP_counterexample}

We have $\setbarU = \{ u \in \Cbb \mid |u| \leq 1 \}$ and $F_\lambda(u) = |u| - \lambda |u|^2$.
It can be verified that for $0 \leq |u| < 1/(2\lambda)$, $F_\lambda(u)$ increases as $|u|$ increases;
specifically, if we let $z= |u|$, and $g(z) = z - \lambda z^2$, we see that $g'(z)= 1 - 2 \lambda z > 0$ for $0 \leq z < 1/(2\lambda)$.
This implies that $F_\lambda(0)  \leq F_\lambda(u)$ for all $u$ such that $|u| \leq 1/(4\lambda)$,
and thus
$u= 0$ is a locally optimal solution to Problem~\eqref{eq:NEP}.
However, $u=0$ is infeasible for Problem~\eqref{eq:NEP_orig}.

\subsection{Proof of Theorem \ref{thm:GEMM_conv}}
\label{app:proof_convergence}
The update \eqref{eq:GEMM_loop} of GEMM can be written as
\[
\bx^{k+1}= \arg\min_{\bx} \frac{\beta_k}{2} \left\|\bx-\left(\bz^k-\frac{1}{\beta_k}\nabla_\bx  G(\bz^k|\bx^k)\right)\right\|^2 + I_{\setX}(\bx).
\]
From the first-order optimality of $\bx^{k+1}$, we have
\[
    \mathbf{0} \in \beta_k(\bx^{k+1}-\bz^k)+\nabla_\bx G(\bz^k|\bx^k) + \partial I_{\setX}(\bx^{k+1}).
\]
Let $\bv^{k+1}\in  \partial I_{\setX}(\bx^{k+1})$ be  such that
\begin{equation}\label{eq:GEMM_opt}
    \mathbf{0} = \beta_k(\bx^{k+1}-\bz^k)+\nabla_\bx G(\bz^k|\bx^k) + \bv^{k+1}.
\end{equation}
Then, we have
\ifconfver
\begin{align}\label{eq:dist}
  & {\rm dist}(\mathbf{0},\nabla F(\bx^{k+1})+ \partial I_{\setX}(\bx^{k+1}))\nonumber \\
   & ~ \leq \| \nabla F(\bx^{k+1})+\bv^{k+1} \| \nonumber \\
     & ~ =  \| \nabla_\bx G(\bz^{k}|\bx^{k}) + \beta_{k} (\bx^{k+1} - \bz^{k})- \nabla F(\bx^{k+1}) \|\nonumber \\
     & ~ \leq \| \nabla_\bx G(\bz^{k}|\bx^{k})- \nabla F(\bx^{k+1})||+ \beta_{k} ||\bx^{k+1} - \bz^{k} \|.
\end{align}
\else
\begin{align}\label{eq:dist}
{\rm dist}(\mathbf{0},\nabla F(\bx^{k+1})+ \partial I_{\setX}(\bx^{k+1}))
	\leq & \| \nabla F(\bx^{k+1})+\bv^{k+1} \| \nonumber \\
	= & \| \nabla_\bx G(\bz^{k}|\bx^{k}) + \beta_{k} (\bx^{k+1} - \bz^{k})- \nabla F(\bx^{k+1}) \|\nonumber \\
	\leq & \| \nabla_\bx G(\bz^{k}|\bx^{k})- \nabla F(\bx^{k+1})||+ \beta_{k} ||\bx^{k+1} - \bz^{k} \|.
\end{align}
\fi
Now, we characterize the two terms in \eqref{eq:dist}. First,
\ifconfver
    \begin{equation}\label{eq:grad_bound}
      \begin{split}
         & \| \nabla_\bx G(\bz^{k}|\bx^{k})- \nabla F(\bx^{k+1}) \| \\
          & ~ = \| \nabla_\bx G(\bz^{k}|\bx^{k}) -\nabla_\bx G(\bx^{k+1}|\bx^{k+1})  \| \\
     & ~ \leq \| \nabla_\bx G(\bz^{k}|\bx^{k}) -\nabla_\bx G(\bx^{k}|\bx^{k}) \|  \\
     & ~ + \| \nabla_\bx G(\bx^{k}|\bx^{k})-\nabla_\bx G(\bx^{k+1}|\bx^{k+1}) \|  \\
     & ~ \leq L_G  \| \bx^{k} - \bz^{k} \| +L_{F} \| \bx^{k}-\bx^{k+1} \| \\
    & ~ = L_G\alpha_k \| \bx^{k} - \bx^{k-1}  \| +L_{F} \|  \bx^{k}-\bx^{k+1} \|,
      \end{split}
    \end{equation}
\else
	\begin{align}
	 \| \nabla_\bx G(\bz^{k}|\bx^{k})- \nabla F(\bx^{k+1}) \|
= & \| \nabla_\bx G(\bz^{k}|\bx^{k}) -\nabla_\bx G(\bx^{k+1}|\bx^{k+1})  \| \nonumber \\
\leq & \| \nabla_\bx G(\bz^{k}|\bx^{k}) -\nabla_\bx G(\bx^{k}|\bx^{k}) \|   + \| \nabla_\bx G(\bx^{k}|\bx^{k})-\nabla_\bx G(\bx^{k+1}|\bx^{k+1}) \|  \nonumber \\
\leq & L_G  \| \bx^{k} - \bz^{k} \| +L_{F} \| \bx^{k}-\bx^{k+1} \| \nonumber  \\
= &  L_G\alpha_k \| \bx^{k} - \bx^{k-1}  \| +L_{F} \|  \bx^{k}-\bx^{k+1} \|,
	\label{eq:grad_bound}
	\end{align}
\fi
    where the first equation is due to $\nabla_\bx G(\bx|\bx)=\nabla F(\bx)$; the third equation is due to the Lipschitz continuity of $\nabla F(\bx)$ and $\nabla_{\bx}G(\bx | \bx^k)$; the fourth equation uses $\bz^{k}=\bx^{k}+ \alpha_{k}(\bx^{k}-\bx^{k-1})$.
Second,
	\begin{equation}\label{eq:point_bound}
	\begin{split}
	\beta_{k} \| \bx^{k+1} - \bz^{k} \|
	= &   \beta_{k} \|  \bx^{k}-\bx^{k+1}+\alpha_k (\bx^k-\bx^{k-1}) \| \\
	\leq &  \alpha_k \beta_k \| \bx^{k} - \bx^{k-1} \| +\beta_k \| \bx^{k}-\bx^{k+1} \|. \\
	\end{split}
	\end{equation}
By combining the results in \eqref{eq:dist}, \eqref{eq:grad_bound} and \eqref{eq:point_bound}, we get
\ifconfver
	\begin{align}\label{eq:sum_bound}
	  %\begin{split}
	  &{\rm dist}(\mathbf{0},\nabla F(\bx^{k+1})+ \partial I_{\setX}(\bx^{k+1})) \nonumber \\
	         &  ~ \leq \alpha_k (L_G+\beta_k) \| \bx^{k} - \bx^{k-1} \|   +(L_{F}+\beta_k) \|   \bx^{k}-\bx^{k+1}  \|  \nonumber \\
%	     & ~ \leq \bar{\alpha}(1+c_2)L_G \|   \bx^{k} - \bx^{k-1} \|  \\
%	    & ~ + (L_F+c_2 L_G)  \|   \bx^{k}-\bx^{k+1} \|   \\
	     & ~ \leq C_1 \left( \|  \bx^{k} - \bx^{k-1} \|  + \|  \bx^{k}-\bx^{k+1} \| \right),
	  %\end{split}
	\end{align}
\else
	\begin{equation}\label{eq:sum_bound}
	\begin{split}
	{\rm dist}(\mathbf{0},\nabla F(\bx^{k+1})+ \partial I_{\setX}(\bx^{k+1}))
	\leq &  \alpha_k (L_G+\beta_k) \| \bx^{k} - \bx^{k-1} \|
	 +(L_{F}+\beta_k) \|   \bx^{k}-\bx^{k+1}  \|  \\
%	\leq &\bar{\alpha}(1+c_2)L_G \|   \bx^{k} - \bx^{k-1} \|   + (L_F+c_2 L_G)  \|   \bx^{k}-\bx^{k+1} \|   \\
	\leq & C_1 \left( \|  \bx^{k} - \bx^{k-1} \|  + \|  \bx^{k}-\bx^{k+1} \| \right),
	\end{split}
	\end{equation}
\fi
where	$C_1 = \max\{ \bar{\alpha}(1+c_2)L_G, L_F+c_2L_G\}$; note that the second and the last equation is due to $\alpha_k \leq \bar{\alpha}$  and $c_1 L_G\leq \beta_k \leq c_2 L_G$.

Next, consider the following lemma.
\begin{Lemma}\label{Lem:exPG}\cite[Lemma 2.2]{Yin2013}
 Let
  \[
  \bx^{+}=\Pi_{\setX} (\bz- \frac{1}{\beta}\nabla H(\bz)),
  \]
  where $\bz=\bx +\alpha (\bx-\bar{\bx})$, $\bx, \bar{\bx}\in \setX$, $\alpha\geq 0$; $H$
   is convex and has Lipschitz continuous gradient; $\setX$ is convex;
    $\beta_k$ is chosen to satisfy
    \[
    H(\bx^+) \leq H(\bz) +\langle \nabla  H(\bz), \bx^{+}-\bz \rangle +\frac{\beta}{2} \| \bx^{+}-\bz \|^2.
    \]
Then, it holds that
    \begin{equation*}
    \begin{split}
   H(\bx)-H(\bx^{+})\geq \frac{\beta}{2}\left( \| \bx^{+}-\bx \|^2-\alpha^2 \| \bx-\bar{\bx} \|^2\right).
    \end{split}
  \end{equation*}
\end{Lemma}

%\medskip

%By leveraging Lemma~\ref{Lem:exPG}, we are ready to show   Theorem \ref{thm:GEMM_conv}.
According to the update rule \eqref{eq:GEMM_loop}--\eqref{eq:desc_exa2}, we have
	\begin{equation*}
	\begin{split}
	F(\bx^{k})-F(\bx^{k+1})
	\geq& G(\bx^{k}|\bx^{k})-G(\bx^{k+1}|\bx^{k}) \\
	\geq & \frac{\beta_k}{2}( \| \bx^{k+1}-\bx^{k} \|^2-\bar{\alpha}^2 \| \bx^{k}-\bx^{k-1} \|^2),
	\end{split}
	\end{equation*}
where the first equation is due to $F(\bx)=G(\bx|\bx)$ and $G(\bx| \bar{\bx})\geq F(\bar{\bx})$;  the second equation is due to Lemma~\ref{Lem:exPG} (with $H=G(\cdot | \bx^{k})$) and $\alpha_k \leq \bar{\alpha}$. As a result, we get
\ifconfver
	\begin{align}\label{eq:run_sum}
	   & F(\bx^{0})-F(\bx^{k+1})\nonumber\\
	& =  \sum_{k'=0}^{k}F(\bx^{k'})-F(\bx^{k'+1})\nonumber\\
	 & \geq \sum_{k'=0}^{k}  \frac{\beta_{k'}}{2}( \| \bx^{k'+1}-\bx^{k'} \|^2-\bar{\alpha}^2 \|\bx^{k'}-\bx^{k'-1}\|^2)\nonumber\\
	   & = \sum_{k'=0}^{k-1} \frac{\beta_{k'} - \bar{\alpha}^2 \beta_{k'+1}}{2} \| \bx^{k'+1} - \bx^{k'}\|^2 + \frac{\beta_k}{2} \| \bx^{k+1} -\bx^k \|^2\nonumber \\
	   & \geq \sum_{k'=0}^{k} \frac{\beta_{k'} - \bar{\alpha}^2 \beta_{k'+1}}{2} \| \bx^{k'+1} - \bx^{k'}\|^2 \nonumber \\
	   & \geq  \sum_{k'=0}^{k}  \frac{c_1 L_G \mu}{2} \| \bx^{k'+1}-\bx^{k'} \|^2,
	\end{align}
\else
	\begin{align}\label{eq:run_sum}
	 F(\bx^{0})-F(\bx^{k+1})
	= &\sum_{k'=0}^{k}F(\bx^{k'})-F(\bx^{k'+1})\nonumber\\
	\geq &\sum_{k'=0}^{k}  \frac{\beta_{k'}}{2}( \| \bx^{k'+1}-\bx^{k'} \|^2-\bar{\alpha}^2 \|\bx^{k'}-\bx^{k'-1}\|^2)\nonumber\\
	=& \sum_{k'=0}^{k-1} \frac{\beta_{k'} - \bar{\alpha}^2 \beta_{k'+1}}{2} \| \bx^{k'+1} - \bx^{k'}\|^2 + \frac{\beta_k}{2} \| \bx^{k+1} -\bx^k \|^2\nonumber \\
	\geq& \sum_{k'=0}^{k} \frac{\beta_{k'} - \bar{\alpha}^2 \beta_{k'+1}}{2} \| \bx^{k'+1} - \bx^{k'}\|^2 \nonumber \\
	\geq &\sum_{k'=0}^{k}  \frac{c_1 L_G \mu}{2} \| \bx^{k'+1}-\bx^{k'} \|^2,
	\end{align}
\fi
where the last inequality is due to $\beta_{k'}\geq c_1 L_G$, $\beta_{k'+1} \leq c_2 L_G$ and $\bar{\alpha}=\sqrt{c_1(1-\mu)/c_2}$.
From \eqref{eq:run_sum}, we get
\ifconfver
	\begin{equation*}
	  \begin{split}
	    &F(\bx^{0})- F^{\star}\\
	     &  \geq F(\bx^{0})- F(\bx^{k+1}) \\
	     & \geq \frac{c_1 L_G \mu}{2} \frac{k}{2} ~\min_{k'=0,\ldots, k} \| \bx^{k'+1}-\bx^{k'} \|^2 +\| \bx^{k'}-\bx^{k'-1} \|^2.
	  \end{split}
	\end{equation*}
\else
	\begin{equation*}
	\begin{split}
	 F(\bx^{0})- F^{\star}
	&  \geq F(\bx^{0})- F(\bx^{k+1}) \\
	& \geq \frac{c_1 L_G \mu}{2} \frac{k}{2} ~\min_{k'=0,\ldots, k} \| \bx^{k'+1}-\bx^{k'} \|^2 +\| \bx^{k'}-\bx^{k'-1} \|^2.
	\end{split}
	\end{equation*}
\fi
By using $a+b \leq \sqrt{2(a^2+b^2)}$, we have
\ifconfver
	\begin{equation}\label{eq:seq_bound}
	\begin{split}
	   & \min_{k'=0,\ldots, k} \| \bx^{k'+1}-\bx^{k'} \|  + \| \bx^{k'}-\bx^{k'-1} \| \\
	     & ~ \leq \sqrt{\frac{8}{c_1 L_G \mu k}(F(\bx^{0})- F^{\star})}.
	   \end{split}
	\end{equation}
\else
	\begin{equation}\label{eq:seq_bound}
	 \min_{k'=0,\ldots, k} \| \bx^{k'+1}-\bx^{k'} \|  + \| \bx^{k'}-\bx^{k'-1} \|
	\leq \sqrt{\frac{8}{c_1 L_G \mu k}(F(\bx^{0})- F^{\star})}.
	\end{equation}
\fi
Substituting \eqref{eq:seq_bound} in \eqref{eq:sum_bound} yields
\ifconfver
	\begin{equation*}
	  \begin{split}
	 &\min_{k'=0,\ldots, k} {\rm dist}(\mathbf{0},\nabla F(\bx^{k'+1})+\partial I_{\setX}(\bx^{k'+1}))\\
	 & ~ \leq  C_1 \sqrt{\frac{8}{c_1 L_G \mu k}(F(\bx^{0})- F^{\star})}.
	  \end{split}
	\end{equation*}
\else
	\begin{equation*}
	\min_{k'=0,\ldots, k} {\rm dist}(\mathbf{0},\nabla F(\bx^{k'+1})+\partial I_{\setX}(\bx^{k'+1}))
	\leq  C_1 \sqrt{\frac{8}{c_1 L_G \mu k}(F(\bx^{0})- F^{\star})}.
	\end{equation*}
\fi
The proof is complete.
%where the second inequality is due to $\max\{ \beta_{k-1},\beta_{k} \} \leq c_2 L_g$.
%It leads to the convergence rate
%\begin{equation*}
% \min_{k=1,\ldots, N} r_{k}+r_{k+1}\leq c_2 L_g  \sqrt{\frac{8}{c_1 L_g (1-\bar{\alpha}^2)N}(f(\bx^{0})- f^{\star})}.
%\end{equation*}
%\hfill $\blacksquare$

%-----------------------------------
%\vfill\pagebreak
%\newpage

\bibliographystyle{IEEEtran}
%\bibliography{refs}

\end{document}